\documentclass[aps,pre,reprint,showpacs,floatfix,groupedaddress,nofootinbib]{revtex4-1}

\usepackage{amscd}
\usepackage{amsmath}
\usepackage{amssymb}
\usepackage{amsthm}
\usepackage{bm}
\usepackage{epsfig}
\usepackage{epstopdf}
\usepackage{graphicx}
\usepackage{latexsym}
\usepackage{mathtools}
\usepackage{mathrsfs}
\usepackage{subfigure}
\usepackage{appendix}
\usepackage{multirow}

\makeatletter
\DeclareRobustCommand*\textsubscript[1]{%
  \@textsubscript{\selectfont#1}}
\def\@textsubscript#1{%
  {\m@th\ensuremath{_{\mbox{\fontsize\sf@size\z@#1}}}}}
\makeatother

\begin{document}

\title{Electron interactions, spin-orbit coupling, intersite correlations in pyrochlore iridates}
\author{Runzhi Wang, Ara Go, Andrew J. Millis}
\affiliation
{
 Department of Physics, Columbia University, New York, New York 10027.
} 
\date{\today}

\begin{abstract}
We perform combined density functional and dynamical mean-field  calculations to study  the pyrochlore iridates Lu$_2$Ir$_2$O$_7$, Y$_2$Ir$_2$O$_7$ and Eu$_2$Ir$_2$O$_7$. Both single-site and cluster dynamical mean-field calculations are performed and spin-orbit coupling is included.  Paramagnetic metallic phases, antiferromagnetic metallic phases with tilted Weyl cones and antiferromagnetic insulating phases are found. The magnetic phases display all-in/all-out magnetic ordering, consistent with previous studies. Unusually for electronically three dimensional materials, the single-site dynamical mean-field approximation fails to reproduce qualitative material trends, predicting in particular that  the paramagnetic phase properties of Y$_2$Ir$_2$O$_7$ and Eu$_2$Ir$_2$O$_7$ are almost identical, although in experiments the Y compound has a much higher resistance than  the Eu compound. This qualitative failure is attributed to the importance of intersite magnetic correlations in the physics of these materials.
\end{abstract}


\maketitle

\section{Introduction}
The interplay of electron-electron interactions  and spin-orbit coupling (SOC) is a central topic in  quantum materials \cite{Pesin2010}. The combination has been predicted to lead to  novel phases including chiral spin liquids \cite{spinliqiud-prl, spinliqiud-nature}, Weyl semimetals (WSM) \cite{dft, tb-HF1, tb-HF2}, and axion insulators (AI) \cite{dft, AI}.  This physics may be particularly relevant to the  iridium oxides, which are characterized by spin orbit  and correlation energies that are comparable to each other and to the conduction bandwidth.  The pyrochlore iridates R$_2$Ir$_2$O$_7$ (R-227, R=rare earth elements or Y) have been intensively studied in this context. 

By means of a ``plus $U$'' extension of density functional theory, Wan et al. predicted that  Y-227 was a Weyl semimetal, with all-in/all-out (AIAO) antiferromagnetic order \cite{dft}. The WSM phase was later found in model systems studies applying the Hartree-Fock \cite{tb-HF1, tb-HF2} and cluster dynamical mean-field \cite{AI} approximation to a tight binding model. The possibility of an axion insulator phase was suggested in Ref. ~\cite{dft} (although the phase was not predicted by the DFT+U used in this reference) and the phase was found by Go et al. \cite{AI} who argued that  both the insulating noninteracting phase and the CDMFT method are necessary to realize this phase.

Experimental studies of the R-227 pyrochlore family of compounds  reveal a systematic dependence of properties on R.  Pr-227 is a paramagnetic metal with a resistivity that decreases as temperature decreases down to  the lowest measured temperature \cite{spinliqiud-prl}, while as R is changed across the rare earth series the resistivity increases and a metal insulator transition (MIT) occurs, with a transition  temperature and optical gap that depends systematically on R \cite{metal-nonmetal-Pr-Nd-Sm-Eu, MIT-Nd-Sm-Eu, MIT-Nd-Sm-Eu-Gd-Tb-Dy-Ho, MIT-mag-Eu1, MIT-mag-Eu2, MIT-mag-Yb, MIT-mag-Nd,Ueda16}. The metal-insulator transitions are accompanied by magnetic transitions, the nature of which is still under debate. Early work by Taira et al. suggested the magnetism was generalically spin-glass-like  \cite{mag-Y-Sm-Eu-Lu}, and this finding  was recently confirmed by Kumar and Pramanik \cite{glass-like}. Onset of spin precession in muon spin rotation experiments indicated the presence of long-range magnetic order in Eu-227, Y-227, Yb-227 and Nd-227  \cite{MIT-mag-Eu1, MIT-mag-Yb, MIT-mag-Nd}, but most neutron scattering measurements did not detect magnetic order  for Nd-227 and Y-227 \cite{MIT-mag-Yb, neutron-Nd, neutron-Y}. Interpretation of the neutron results is however complicated by the relatively small values of the ordered moment and the large neutron absorption cross section of Ir. An intermediate disordered phase between the magnetic transition and the onset of the long-range order has been reported in Nd-227, Sm-227 and Y-227 \cite{MIT-mag-Yb, neutron-Nd, intermediate-phase-Nd-Sm}, but most studies agree that  long-range order appears concomitantly with the magnetic transition in Eu-227  \cite{MIT-mag-Eu1, MIT-mag-Eu2}; Yb-227  \cite{MIT-mag-Yb} and Nd-227  \cite{MIT-mag-Nd}. Strong evidence of  AIAO magnetic order was reported in recent studies \cite{AIAO-Nd, AIAO-Eu, AIAO-Y-Eu-Nd} and  for Nd-227 a direct determination of the AIAO magnetic structure was reported by the neutron scattering measurement  \cite{Determination-AIAO-Nd2Ir2O7}.
An AIAO-type structure, which breaks the time-reversal symmetry while preserving the inversion symmetry, is essential for the realization of the theoretically predicted Weyl semimetal  phase in pyrochlore iridates, but whether or not this phase occurs remains unclear. A recent optical experiment gave indications for a WSM in Eu-227 \cite{WSM-optical-Eu} and in Ref. ~\cite{Ueda16} optical data were interpreted as indicating to the presence of a Weyl semimetal state in Sm-227 and perhaps Eu-227, while angle-resolved photoemission spectroscopy (ARPES) measurements in Nd-227 failed to observe the Weyl points \cite{WSM-ARPES-Nd}.

Motivated by the theoretical predictions of  novel phases and discrepancies between experimental reports, two density functional plus single-site dynamical mean-field (DFT+sDMFT) studies of pyrochlore iridates were recently carried out. In Ref. ~\cite{lda-dmft-Y}, Shinaoka et al. obtained a phase diagram for Y-227 in correlation strength ($U$)-temperature ($T$) plane. As $U$ was varied they found a crossover between a paramagnetic metallic phase and paramagnetic insulating phase at high $T$, and a coupled MIT and magnetic transition to an AIAO insulator at lower $T$. A  WSM phase was not found. Zhang et al. \cite{lda-dmft-R-dep} studied the R-dependence of the MIT  by performing DFT+sDMFT calculations for several members of the family.  They found good agreement with the measured trends of magnetic transition temperature with R, and reported that for the $U$ value they considered all non-magnetic phases were metallic.  The zero temperature metal-insulator  boundary  was predicted to be between Nd-227 and Pr-227 and consistently with Shinaoka et al the magnetic phases were topologically trivial insulators: a Weyl semimetal phase was not found. 

In this work, we go beyond previous work by performing  DFT plus cluster DMFT (DFT+CDMFT) calculations for the pyrochlore iridates. To clarify the direct contribution from the Ir sublattice, we focus on the compounds Lu-227, Y-227 and Eu-227 for which the R ion is nonmagnetic. A key finding of the cluster calculations is a relatively wide range of stability of a Weyl metal phase with tilted Weyl cones, suggesting a  key role played by  intersite quantum fluctuations in stabilizing topological physics. 

We also consider the dependence of physical properties on rare earth ion R and perform a critical comparison of the predictions of single-site and cluster dynamical mean-field theory.  We find that the single-site approximation overpredicts gap values  and fails to account for the substantial difference in observed properties of the Y and Eu compounds, predicting instead that the behavior of these two materials should be essentially identical at temperatures greater than the N\'eel temperature. It also predicts that the insulating gap in these compounds is much larger than any experimentally reasonable value. These failures of the single-site dynamical mean-field approximation are not expected in electronically three dimensional materials. On the other hand, our CDMFT calculations  yield a reasonable gap range which is consistent with the transport data, and, when magnetic order is included, account for a considerable portion of the difference between the Y and Eu materials. We demonstrate that it is the intersite correlations, most likely of antiferromagnetic origin, that leads to the material-dependence in the physics of Y-227 and Eu-227.   

The reminder of this paper is organized as follows: In Sec. \ref{methods}, we summarize the techniques used in our calculations; In Sec. \ref{phase diagram}, we illustrate the generic phase diagram obtained by means of CDMFT; In Sec. \ref{MIT}, we take Eu-227, Y-227 and Lu-227 as examples to discuss the R-dependent MIT by comparing the results within DFT+U+SO and DFT+sDMFT/CDMFT in both paramagnetic and magnetic states.  

\section{Methods} \label{methods}

\begin{figure}
\begin{center} 
\subfigure[crystal structure]{
\label{structure}   
\includegraphics[width=0.4\linewidth]{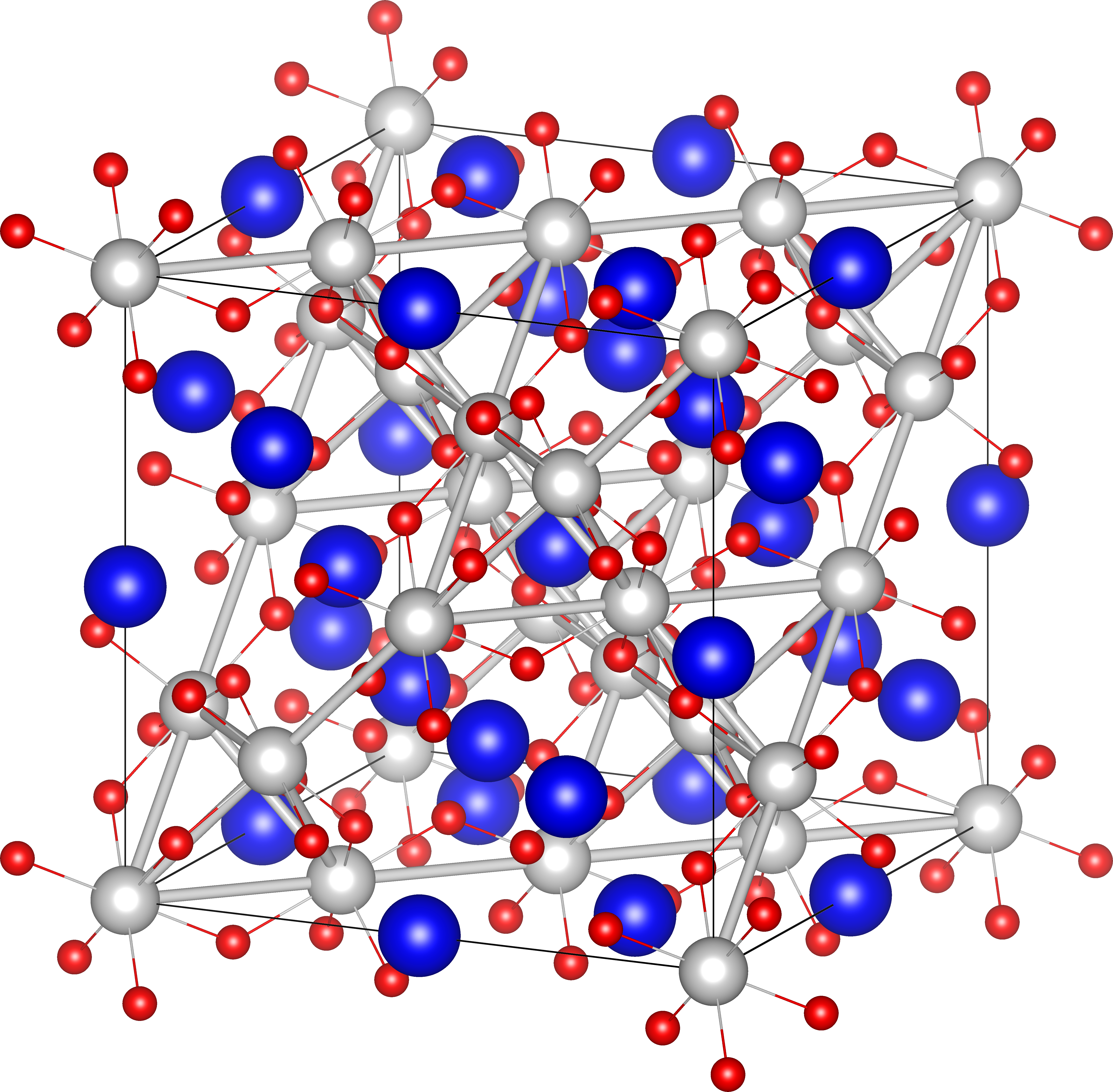}}
\subfigure[band structure]{
\label{band}   
\includegraphics[width=0.56\linewidth]{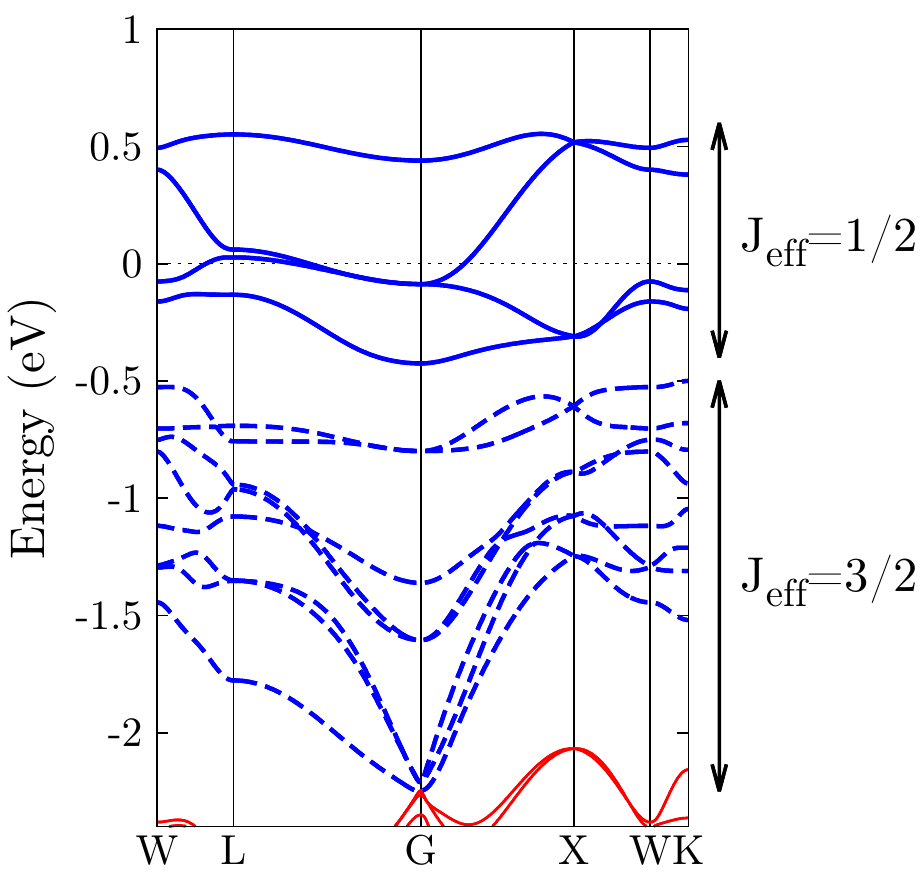}}
\subfigure[Brillouin zone]{
\label{BZ}   
\includegraphics[width=0.8\linewidth, angle=-90]{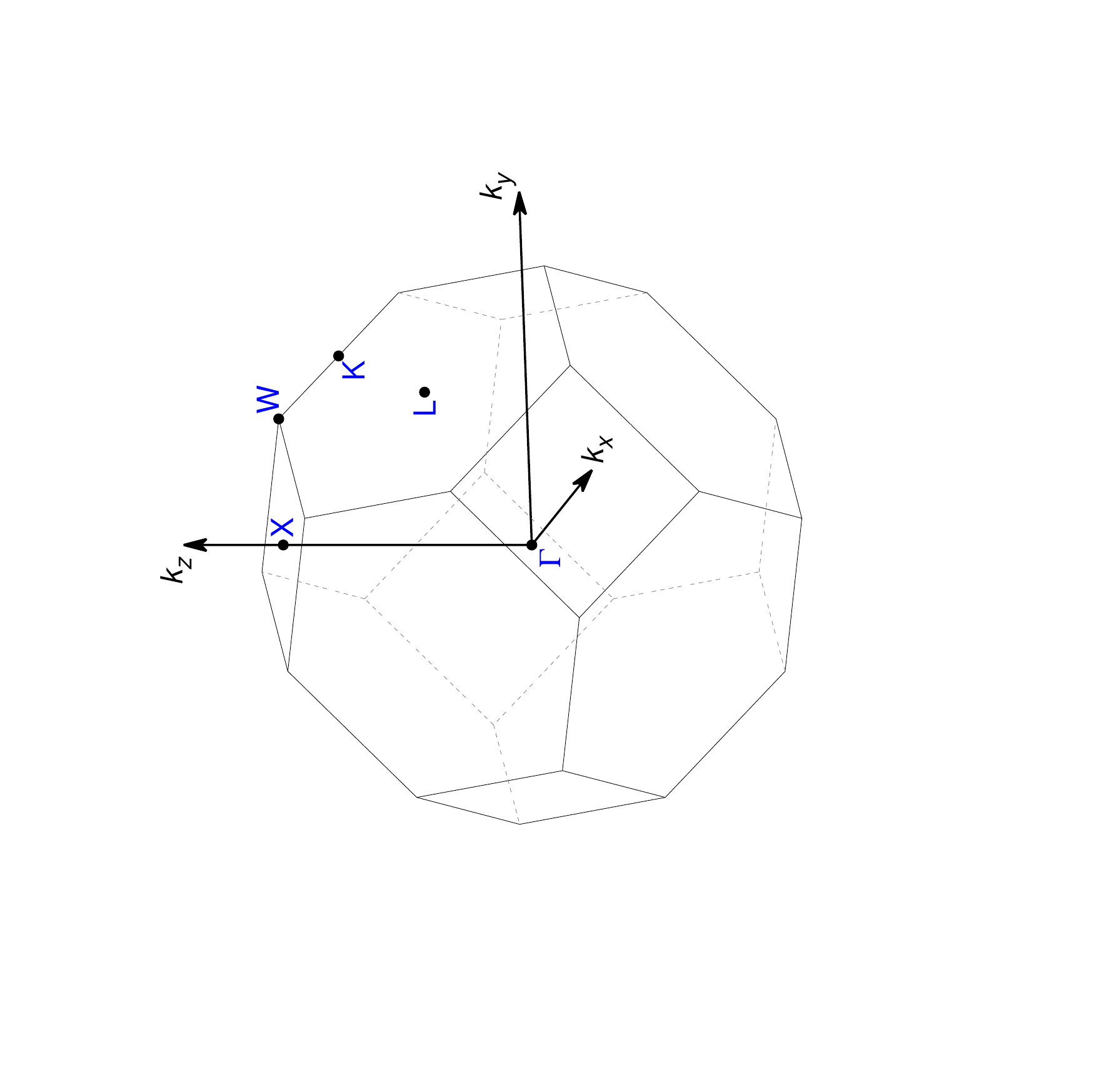}}
\caption{(a) Crystal structure of pyrochlore iridates. The large dark sphere (blue online) is the R site; the large light sphere (grey online) is Ir; the small one (red online) is O. (b) Band structure of Y-227 within paramagnetic DFT+U+SO. The energy window is chosen to include Ir $t_{2g}$ bands, which split into the higher-energy $J_\mathrm{eff}=1/2$ (dark (blue online) solid lines) and lower-energy $J_\mathrm{eff}=3/2$ (dark (blue online) dashed lines) manifold in the presence of SOC. The light (red online) bands are irrelevant non-iridium d-states. (c) Brillouin zone of the fcc lattice.
}  
\label{structure-band}
\end{center}
\end{figure}

The structure of pyrochlore iridates is presented in Fig. \ref{structure}. The Ir atoms form  corner-sharing tetrahedra and each Ir is surrounded by a distorted oxygen octahedron which determines the local symmetry of the Ir site.  The relevant electrons reside in the Ir 5d shell and are subject to both on-site correlations and strong spin-orbit coupling. We treat this physics using the density functional plus dynamical mean-field approximation. As a first step, we employ the Vienna Ab-initio Simulation Package (VASP) \cite{vasp1, vasp2, vasp3, vasp4, vasp5}, which is based on projector augmented wave (PAW) method \cite{PAW}, to perform fully relativistic paramagnetic DFT calculations including the on site Coulomb interaction $U$ (within the DFT+U approximation). In our study, we focus on Lu-227, Y-227 and Eu-227 in which the rare earth element is non-magnetic and use the experimental structures \cite{mag-Y-Sm-Eu-Lu} in all calculations. The generalized gradient approximation (GGA) in the Perdew, Burke, and Ernzerfhof (PBE) \cite{PBE} parametrization is used as the exchange-correlation functional. In all calculations, we take $8 \times 8 \times 8$  $k$-point mesh and a plane wave energy cut-off of 500eV. For the DFT+U calculations, 2eV is applied on Ir as the effective on site Coulomb interaction.

Representative results are shown in Fig. \ref{band}. The bands near the fermi level are mainly contributed by Ir $t_{2g}$ states and some mixture with O $p$ states. The strong spin orbit coupling splits the Ir  $t_{2g}$ states into a lower-lying $J_\mathrm{eff}=3/2$ manifold and the higher-energy $J_\mathrm{eff}=1/2$ manifold. The ${4+}$ Ir formal valence leaves the $J_\mathrm{eff}=3/2$ manifold fully filled and electrically inert while   the $J_\mathrm{eff}=1/2$-derived bands are half-filled and  most relevant to the low-energy physics. We use maximally localized Wannier function (MLWF) methods \cite{MLWF}, as implemented in the Wannier90 code \cite{Wannier90} with the VASP interface, to project  the $J_\mathrm{eff}=1/2$ band complex onto a basis of states localized on the Ir atoms. The Wannier bands provide an essentially perfect fit to the calculated DFT bands, as expected since the bands being fit are isolated from other bands by energy gaps.  After the construction, the MLWFs centered on a given site  are rotated to an orientation adapted to the local octahedra. The rotation was determined by comparing the tight-binding Hamiltonian in Ref. ~\cite{AI} to the corresponding Wannier Hamiltonian so that nearest neighbor hopping parameters are the same. It is important to note that the degree to which the $J_\mathrm{eff}=1/2$  and $J_\mathrm{eff}=3/2$ band complexes are separated depends on $U$.  It is also important to note that the low point symmetry of the individual Ir ion and the fact that the on-site angular momentum quantum number is not conserved by the intersite terms in the Hamiltonian mean that the bands labelled as $J_\mathrm{eff}=1/2$  have contributions also from atomic $J_\mathrm{eff}=3/2$ states as well as oxygen atoms. This subtlety is not important for our calculations, which are based on the Wannier fitting to the calculated bands. In the Wannier projection, the spin-orbit coupling effects appear as a spin dependence of the hopping terms.  

We then take the Wannier projected Hamiltonian, add an on-site $U$ to each Ir site, and solve the resulting model using either single-site (sDMFT) or cluster (CDMFT) approximations \cite{dmft, CDMFT1, CDMFT2, cluster}.  In sDMFT calculations, we treat each Ir atom independently, assuming a self energy with only site-local components.   In  our CDMFT calculations we choose the real-space cluster as the four Ir ions on the vertices of a tetrahedron.  We study both paramagnetic and antiferromagnetic phases. In the paramagnetic calculations, bath parameters are constrained in the way determined by the double group analysis for this system, following the method put forward by Koch et al. \cite{group-theory}. 

The CDMFT approximation does not treat all bonds in the physical lattice equally. The self energy computed in the impurity model  includes intersite matrix elements of the self energy among sites in the cluster, but in the impurity model stage the self energy matrix elements involving symmetry-equivalent bonds connecting sites in different clusters are not included. 
In the literature, the process of restoring translational symmetry is referred to as `periodization'. In our case the tetrahedral cluster is a primitive unit cell of the pyrochlore lattice, so the impurity model self energy is periodic, but it does not obey all of the point symmetries of the lattice. Even though the symmetrization matters for the precise location of the Weyl crossing, it makes only a very small difference to the quantities of interest in this paper. Therefore the results presented here are obtained directly from the computed CDMFT self energy. Symmetrization and its consequences (particularly for the behavior near the Weyl points) will be presented in a separate paper.

We use exact diagonalization (ED)  \cite{ED} as the impurity solver.  In the ED method, one approximates the bath by a finite number of orbitals, and errors due to finite bath size are an issue. We perform  cluster DMFT calculations using four correlated and eight bath orbitals ($N_b=8$), which is the largest number we can access, and discuss convergence in an Appendix.  At finite bath size, the DMFT self-consistency is not perfect, so that results for local quanities calculated using the impurity model may differ from those calculated using the lattice Green function.  In this paper, all the physical properties are computed based on the lattice Green function.

\section{Phase diagram, spectral function and magnetic properties} \label{phase diagram}

\begin{figure}[t]
\begin{center}
\includegraphics[width=0.9\columnwidth]{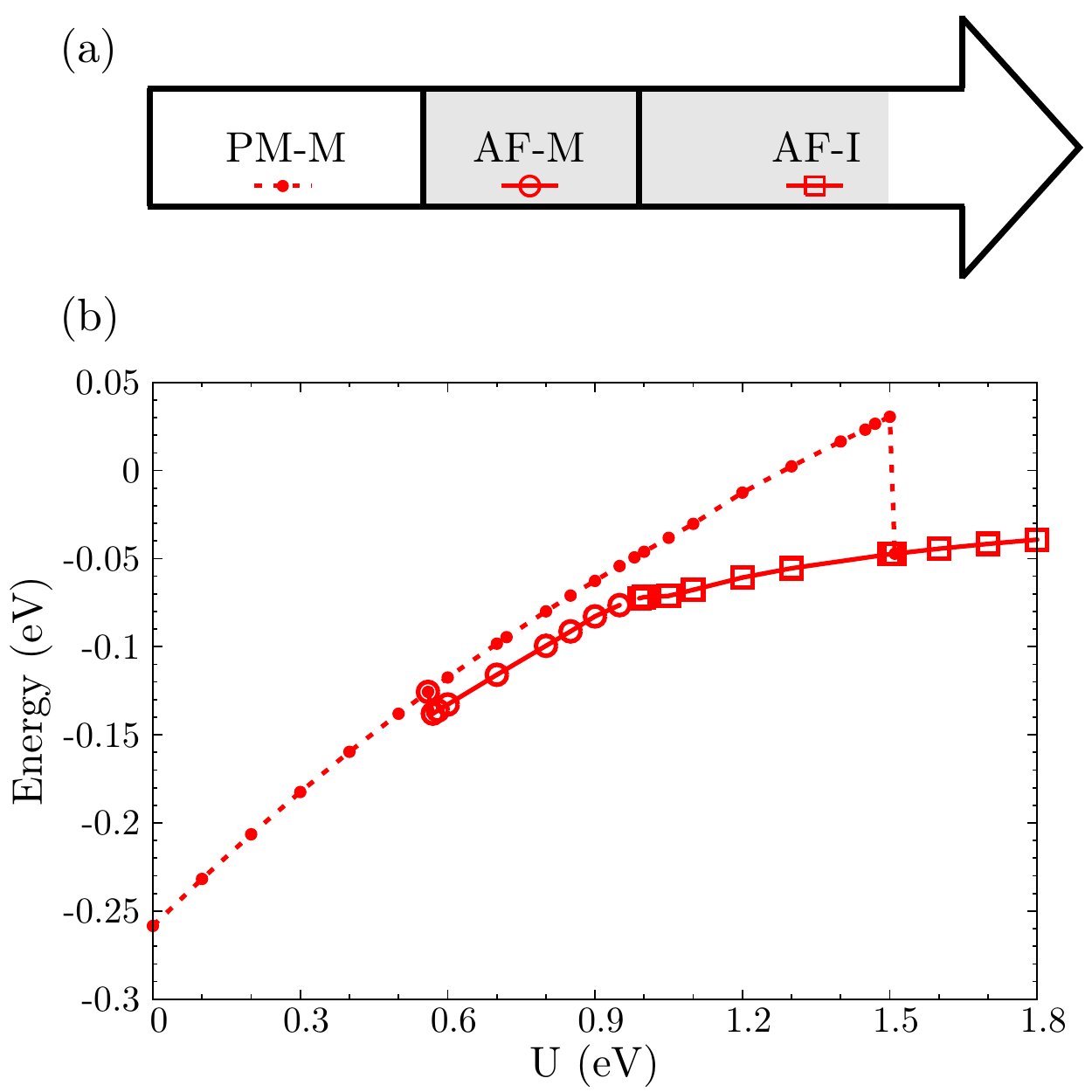}
\caption{(a) Generic ground state phase diagram for pyrochlore iridates obtained with DFT+CDMFT calculations as described in the main text.  The shaded region indicates the interaction strengths for which  the paramagnetic phases are locally stable, but higher in energy compared than the antiferromagnetic phases. Abbreviations: PM-M, paramagnetic metal; AF-M, antiferromagnetic metal with AIAO magnetic ordering and Weyl cones; AF-I, antiferromagnetic insulator with AIAO ordering but without Weyl cones. (b) Energy as a function of the interaction strength for all the states obtained with DFT+CDMFT for Y-227. The small filled circles indicate the energy of the paramagnetic phase. The open squares represent the antiferromagnetic insulating phase and the open circles represent the antiferromagnetic metallic phase. The precise behavior near  $U\sim  0.5-0.6$eV is not resolved due to finite bath size effects.}
\label{phase diagram-gap-mag}
\end{center}
\end{figure}

This section analyses the generic ground-state phase diagram as a function of interaction strength $U$. For definiteness we present results obtained for the Y compound but note that all compounds yield qualitatively similar results, with the only differences being the values of $U$ at which phase transitions occur and the quantitative values of gaps and magnetic moments.  The material dependence will be considered in more detail in the next section. 

We begin with the generic ground state phase diagram obtained from DFT+CDMFT calculations shown in panel (a) of Fig. \ref{phase diagram-gap-mag}. As $U$ is increased we find a  transition from a paramagnetic metal state to an antiferromagnetic metal state with AIAO order and Weyl cones, followed by a metal-insulator transition at larger $U$. The symmetry of the magnetic state does not change across the metal-insulator  transition. Everywhere that it can be stabilized, the antiferromagnetic phase is the ground state, but the paramagnetic phase exists as a higher energy metastable phase over the remarkably wide range of $U$ shown as the shaded region in panel (a).

The  sequence of phases shown in Fig. \ref{phase diagram-gap-mag} (a)  was reported in the DFT+U calculations of Wan et al \cite{dft} and also  in the Hartree-Fock phase diagram\cite{tb-HF1, tb-HF2}.  The intermediate Weyl metal  phase is not found in recent LDA+single-site DMFT studies \cite{lda-dmft-Y, lda-dmft-R-dep}, which report instead a direct transition from paramagnetic metal to AIAO antiferromagnetic insulator.  The CDMFT calculations predict a much wider range of Weyl-metal behavior than that in the Hartree-Fock calculations,  indicating the importance of the intersite quantum fluctuations captured by the CDMFT method. 

\begin{figure}
\begin{center}
\includegraphics[width=0.9\columnwidth]{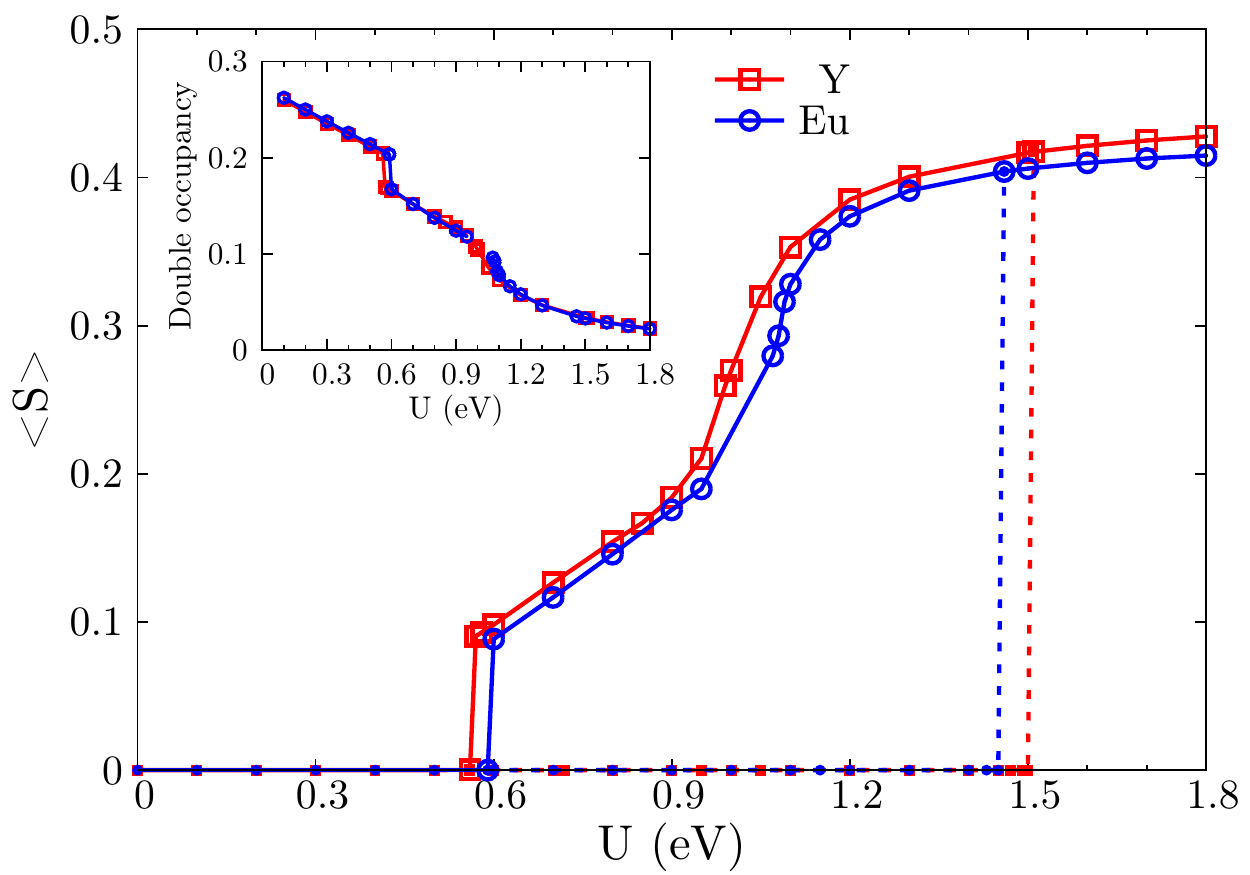}
\caption{$\langle S \rangle$ as a function of the interaction strength for both Y-227 and Eu-227. The results are obtained within CDMFT. The solid lines identifies the evolution of the ground state while the dashed lines identifies the evolution of the metastable phases. The open symbols represent the antiferromagnetic phases which are obtained by decreasing $U$. The filled symbols represent the paramagnetic phases which are obtained by increasing $U$. The inset of this figure shows the $U$-dependent behavior of the double occupancy for the ground state, which is computed as $\frac{1}{U}\frac{1}{\beta}\sum_{n}\sum_{k}\frac{1}{N_c}Tr\frac{1}{2}\mathbf \Sigma(i\omega_{n})\mathbf G(k,i\omega_{n})$. We will leave the discussion of the material dependence to the next section.
}
\label{AF-mag-u}
\end{center}
\end{figure}

Panel (b)  of Fig. \ref{phase diagram-gap-mag} presents the evolution with $U$ of the ground state energy (per correlated orbital), computed from \cite{cdmft-model}:
\begin{eqnarray}
E &=& \frac{1}{\beta}\sum_{n}\sum_{k}\frac{1}{N_c}Tr\left\lbrace\left[\mathbf H_0(k)+\frac{1}{2}\mathbf \Sigma(i\omega_{n})\right]\mathbf G(k,i\omega_{n})\right\rbrace \nonumber \\
 &\label{whatever}
\end{eqnarray}
where $\mathbf H_0$ is the projection of the DFT Hamiltonian on the $J_\mathrm{eff}=1/2$ MLWF basis, $\mathbf G(k,i\omega_{n})$ is the interacting Green function defined as $\mathbf G(k,i\omega_{n})=[(i\omega_{n}+\mu)\mathbf 1-\mathbf H_0(k)-\mathbf \Sigma(i\omega_{n})]^{-1}$ and $\mathbf G$ and $\mathbf \Sigma$ are matrices in the $8\times 8$ space of $J_\mathrm{eff}=1/2$ orbitals on the tetrahedron.  The small discontinuity in energies at the lower $U$ end of the antiferromagnetic phase is a finite bath size error and is related to first order nature of the PM-AF transition in our approximation.  Where it can be stabilized we also report the energy of the metastable paramagnetic metallic state.

To further characterize the phase behavior we calculated the expectation value of the spin operator on each site, $\langle {\vec S}\rangle$. The magnetic order is found to be of the AIAO type.  We present in Fig. \ref{AF-mag-u} the evolution with interaction strength of the size of the moment $\sqrt{\langle S_x \rangle^2+\langle S_y \rangle^2+\langle S_z \rangle^2}$. The magnetic moment values saturate at large $U$. The saturation value is less than the classical limit $\sqrt{\frac{1}{2}(\frac{1}{2}+1)}$,  reflecting quantum fluctuations in the pyrchlore lattice, which are captured by the CDMFT methodology. 

We see that the paramagnetic metal  to antiferromagnetic metal transition is associated with a large jump in moment size. We therefore identify this transition as first order, consistent with the wide hysteresis region where both paramagnetic and antiferromagnetic solutions may be stabilized. The order of the  Weyl metal to trivial insulator transition is less clear. The $E(U)$ curve shown in Fig. ~\ref{phase diagram-gap-mag} indicates in slope of the energy vs $U$ curve, suggesting a first order transition, but gaps and moments seem to vary rapidly, but continuously, suggesting a second order transition. Whether the transition is second order or weakly first order remains to be determined.   

\begin{figure}
\begin{center}
\includegraphics[width=1\columnwidth]{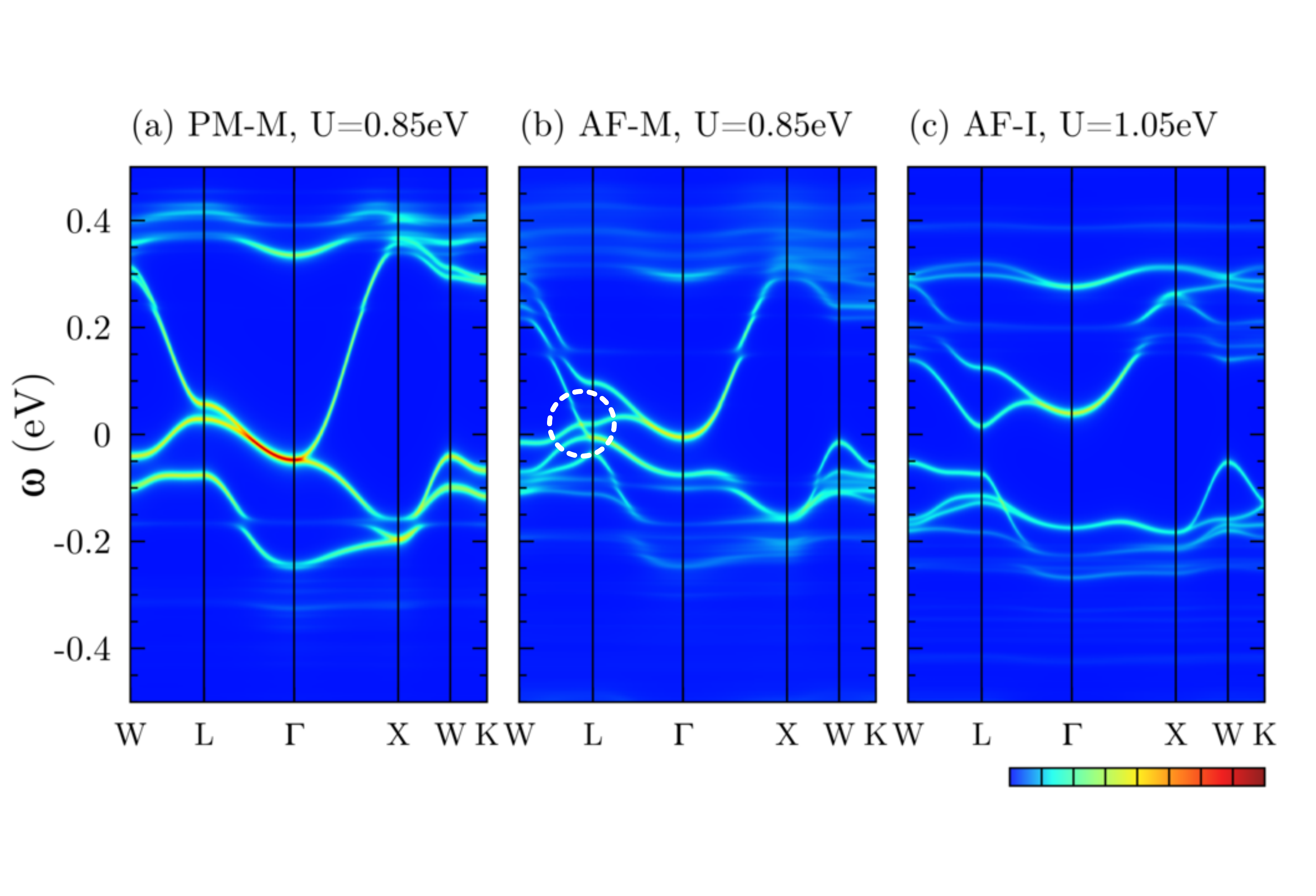}
\caption{Spectral function for Y-227 in (a) paramagnetic metallic phase (b) antiferromagnetic metallic phase and (c) antiferromagnetic insulating phase. The dashed circle of panel (b) highlights the region where the Weyl crossing occurs. The zero of energy is the chemical potential.
}
\label{spectral}
\end{center}
\end{figure}

In Fig. \ref{spectral} we present the DFT+CDMFT electron spectral function (imaginary part of lattice Green function)  for values of interaction strength in the paramagnetic metal (a), antiferromagnetic metal (b) and antiferromagnetic insulator (c) phases. Panel (a) shows that the metastable  paramagnetic metallic phase is a renormalized metal. The circle in panel (b) highlights the band crossing that produces  Weyl cones. The presence of this crossing leads to  metallic pockets at the fermi level. The metal-insulator transition occurring as $U$ is further increased is driven by the combination of  a transition out of the Weyl phase (annihilation of pairs of Weyl points), so the degeneracy at the Weyl point is lifted and a band gap is opened, and by a lifting of the bands at $\Gamma$ point, as shown in Fig. \ref{spectral} (c). The details of the behavior of the Weyl points are complicated, especially near the large-$U$ end point of the antiferromagnetic metallic phase, and  will be discussed in a separate paper.

\section{Dependence of properties on rare earth ion} \label{MIT}

Varying R across the lanthanide rare earth series drives a significant change in properties of the pyrochlore iridates. Pr$_2$Ir$_2$O$_7$ is metallic down to lowest temperature  \cite{spinliqiud-prl} and the materials become progressively more insulating as Pr is replaced by Sm, Eu, Y and Lu \cite{metal-nonmetal-Pr-Nd-Sm-Eu, MIT-Nd-Sm-Eu, MIT-Nd-Sm-Eu-Gd-Tb-Dy-Ho, MIT-mag-Eu1, MIT-mag-Eu2, MIT-mag-Yb, MIT-mag-Nd, review,Ueda16} (While Y is not a lanthanide rare earth the Y compound fits naturally in this progression). The change in behavior is manifest both as an R dependence of the magnetic transition temperature and as a progressive increase in the  magnitude of the high temperature (paramagnetic phase) resistivity (except the Dy compound is reported to be more resistive than the Ho compound \cite{MIT-Nd-Sm-Eu-Gd-Tb-Dy-Ho}). The DFT+single-site DMFT methodology was found by  Zhang, Haule and Vanderbilt \cite{lda-dmft-R-dep} to account quantitatively for the variation with R of the magnetic transition temperature. For the materials studied the variation was traced to a change in bandwidth which was reflected in changes in the hybridization function and for the interaction strength used by Zhang et al at which the paramagnetic phases were stated to be metals.

While all experimental papers report similar trends, the details of the reported resistivities vary, perhaps in part because the properties are very sensitive to the stoichiometry. The consensus is that the Y compound is a paramagnetic insulator, with high temperature resistivity that is large and increases as $T$ is decreased, while the high temperature behaviors of the Eu compound are inconsistent. We fit the reported \cite {metal-nonmetal-Pr-Nd-Sm-Eu, MIT-Nd-Sm-Eu, MIT-Nd-Sm-Eu-Gd-Tb-Dy-Ho, MIT-mag-Eu2, MIT-mag-Yb, AIAO-Eu, transport-Y, transport-Eu, glass-like} resistivities to $\rho(T) =\rho_0 e^{E_g/T}$.  For Eu-227, the analysis let to estimated gap values $E_g=0\sim26$meV; for Y, $E_g=13\sim64$meV. Optical data \cite{Ueda16} show a hard gap of 0.2eV (Eu-227) and 0.4eV (Y-227) at low temperatures, and somewhat smaller and rather broadened gaps in the same materials at room temperature. On the other hand, the high temperature resistivity reported for Y-227 is about tenfold higher than Eu-227 \cite{metal-nonmetal-Pr-Nd-Sm-Eu, Ueda16}.

\begin{figure}
\begin{center}
\includegraphics[width=0.9\columnwidth]{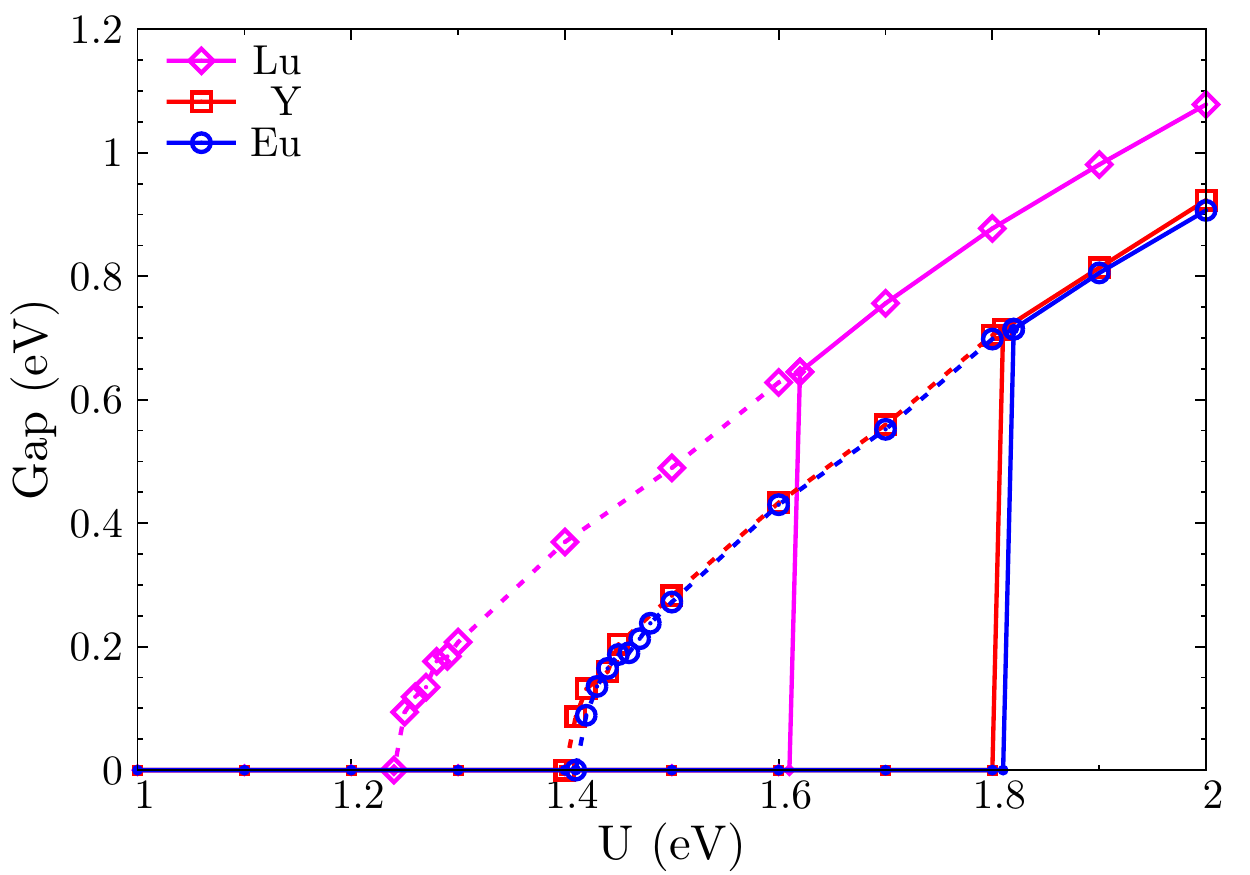}
\caption{Gap size as a function of the interaction strength obtained within paramagnetic sDMFT. For all compounds, the filled symbols represent the metallic phase which is obtained by increasing $U$ while the open symbols represent the insulating phase which is obtained by decreasing $U$. The solid lines identify the evolution of the ground state while the dashed lines represent the metastable phases.}
\label{SDMFT-gap}
\end{center}
\end{figure}

We  first use the single-site dynamical mean-field method to study the $U$-dependence of the spectral gap of Lu-227, Y-227 and Eu-227. We constrain the calculation to the paramagnetic phase. Results are shown   in Fig. \ref{SDMFT-gap}. We see an insulating (non-zero-gap) phase at larger $U$ and a metallic (zero gap) phase at smaller $U$.  As is typically found in the single-site dynamical mean-field method there is a wide coexistence regime where both insulating and metallic phases are locally stable. As is also typical, computations of the energy show that the metallic phase has the lower energy over essentially all of the  coexistence region (some uncertainty remains as to the relative energies of the two phases for $U$ close to the upper boundary of the coexistence region, where the energy difference is comparable with the fitting error resulted from finite number of bath orbitals). The minimum gap magnitude  $\sim 0.7$eV of the globally stable paramagnetic  insulating state is much larger than any gap estimated from transport or optical \cite{Ueda16} data. Within the single-site DMFT methodology, obtaining a globally stable small-gap solution  requires antiferromagnetic order and the consequence is that when the order is melted metallic behavior would occur (see e.g. Ref. ~\cite{lda-dmft-R-dep}). The  moderate gap paramagnetic insulator phase observed in some pyrochlore iridates is  therefore  beyond the scope of the single-site DMFT method. However, it is important to stress that the DFT+single-site DMFT methodology   successfully explains important aspects of the material dependence, as previously shown by Zhang et al ~\cite{lda-dmft-R-dep}. In particular, at any $U$ value, Lu-227 exhibits a noticeably larger gap compared with Y-227 or Eu-227, consistent with experiments. Interestingly, however,  the calculated gaps for Y-227 and Eu-227 are almost identical, in constrast to the obvious material dependence evidence in the paramagnetic phase resistivity.

\begin{figure}
\begin{center}
\includegraphics[width=0.9\columnwidth]{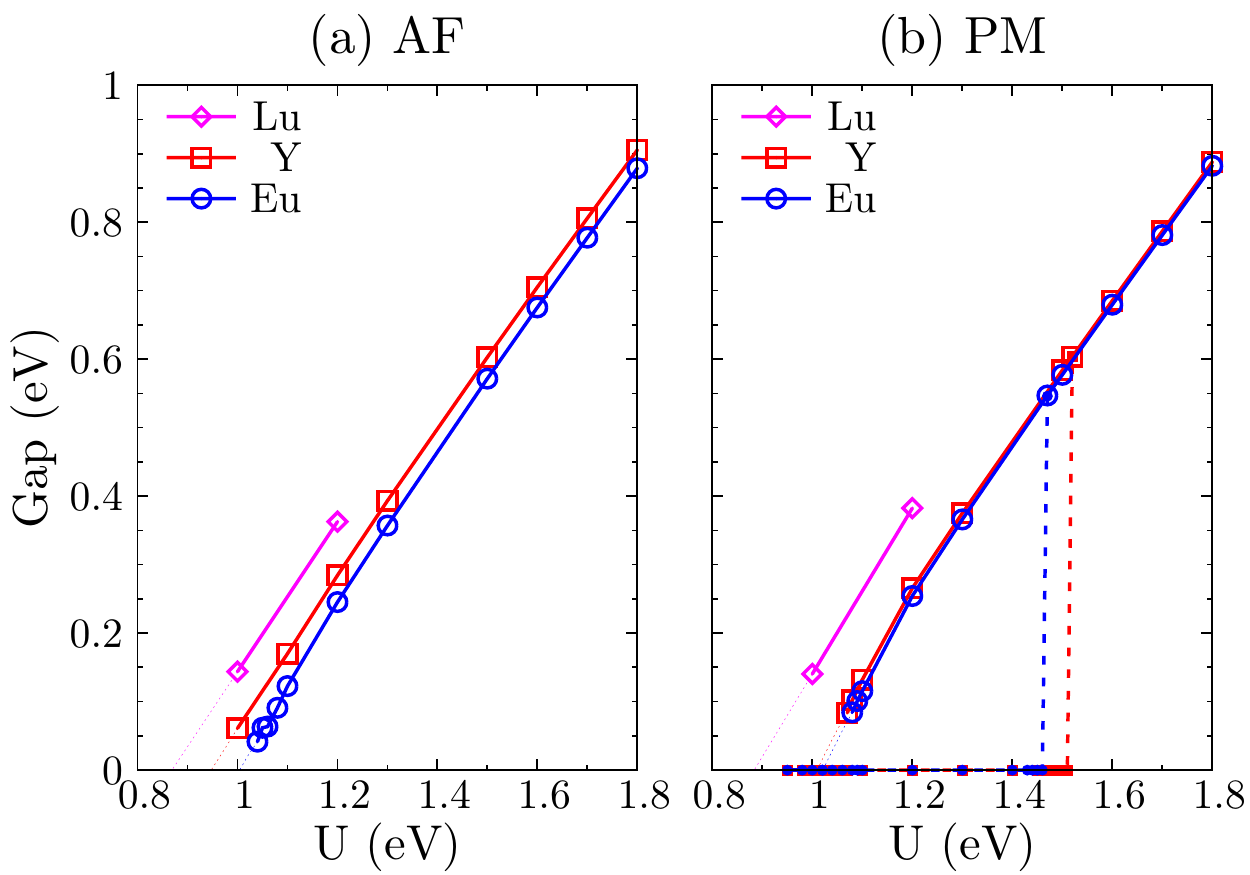}
\caption{Gap size as a function of the interaction strength obtained within (a) magnetic CDMFT (6 bath orbitals) and (b) paramagnetic CDMFT (8 bath orbitals). Convergence difficulties related to the finite bath size prevent us determining the gap size in the vicinity of the metal-insulator transition; the critical  $U$ is estimated using a linear extrapolation (dotted lines). The two panels share the same ordinate. In panel (a), the insulating phase is shown by open symbols; in panel (b), the filled symbols represent the metallic phase which is obtained by increasing $U$ while the open symbols represent the insulating phase which is obtained by decreasing $U$. The solid lines identify the  ground state while the dashed lines represent  metastable phases. 
}
\label{PM-AF-cluster-gap}
\end{center}
\end{figure}

We now turn to the CMDFT calculations, shown in Fig.~\ref{PM-AF-cluster-gap}. Panel (b) shows CDMFT calculations restricted to the paramagnetic phase. As in the single-site approximation we find a small $U$ metallic phase and a larger $U$ insulating phase, with an intermediate $U$ coexistence region and we find that the Lu material has a noticeably larger gap than the Y and Eu materials. In contrast to the single-site approximation it is the insulating state which is stable over the entire coexistence region, thus the CDMFT critical $U$ for the paramagnetic metal-paramagnetic insulator transition is about 60\% of the single-site DMFT critical $U$. A similar behavior and similar contrast in critical $U$ was observed in studies of the two dimensional Hubbard model \cite{Park08,Gull08}. We have not determined whether the paramagnetic metal to paramagnetic insulator transition is second order or weakly first order, but solutions with a very small gap can be sustained. These calculations suggest that intersite correlations are essential in describing the physics of the pyrochlore iridates. 

Panel (b) of Fig.~\ref{PM-AF-cluster-gap} shows that even the paramagnetic phase CDMFT calculations predict essentially no difference between the Y and Eu materials. However,  panel (a) of Fig.~\ref{PM-AF-cluster-gap} shows that a moderate material difference does appear if the antiferromagnetic phase is considered (a material dependence occurs also in the antiferromagnetic phase single-site DMFT calculations, but these calculations are much more difficult to stabilize). The ratio of gap sizes is of course largest near the end point of the insulating phase. The difference, although notable, is rather less than the factor of two reported by Ueda et al in a recent optical study \cite{Ueda16}. By a direct comparison to the optical gaps, we estimate a correlation strength ($U$) of 1.3eV in Y-227 and 1.2eV in Eu-227. Fig. ~\ref{AF-mag-u} shows the material dependence of the magnetization; again the difference is largest in the range $U\sim 0.9-1.1$eV near the metal-insulator transition point. The nearly $10\%$ relative difference in $\langle S \rangle$ for $U$ in this range is consistent with the $\mu$SR measurement \cite{AIAO-Y-Eu-Nd}, in which the local magnetic field was reported to be $\sim10\%$ smaller in Eu-227 than in Y-227.  Interestingly, the double occupancy shown in the inset of Fig. \ref{AF-mag-u} is the same for both materials, indicating that the effective correlation strength is the same. Thus we conclude that in the calculation the difference between the Y and Eu materials arises from an intersite effect related to magnetic ordering (or correlations) and is not directly related to bandwidth or effective correlation strength. 

\begin{figure}[b]
\begin{center}
\includegraphics[width=0.9\columnwidth]{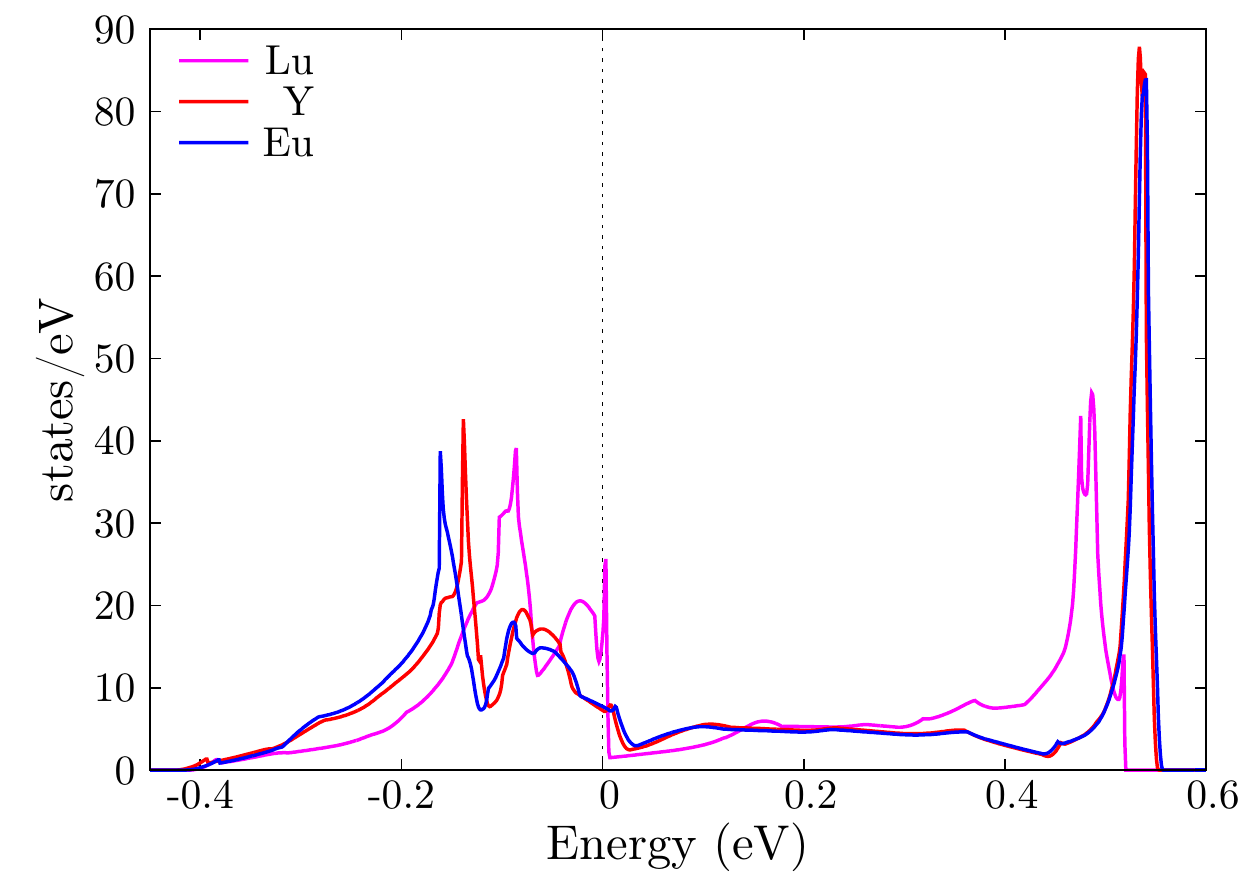}
\caption{Total density of states in $J_\mathrm{eff}=1/2$ manifold obtained within paramagnetic DFT+U+SO calculations for Lu-227, Y-227 and Eu-227. The dashed line denotes the fermi energy.}
\label{dos}
\end{center}
\end{figure}

As the first step towards understanding the origin of the material differences, we present in Fig.~\ref{dos} the total density of states in the paramagnetic state for Lu-227, Y-227 and Eu-227, in an  energy window near the fermi level where the states  arise from  Ir $J_\mathrm{eff}=1/2$ orbitals with some admixture of O $p$ orbitals, by means of DFT+U+SO. We see that Lu-227 has a narrower bandwidth than Y-227 or Eu-227 so that the ratio of correlation strength to bandwidth is larger in the Lu material than in the Y or Eu materials. We conclude, in agreement with Zhang et al \cite{lda-dmft-R-dep}, that the more insulating behavior of Lu-227 can be understood as a bandwidth effect that  is captured by the single-site DMFT approximation.   However, we see that the  density of states of Y-227 and Eu-227 are very similar,  except for a shift of the peak at $E\sim -0.15$eV, supporting the hypothesis that the Y-Eu material difference does not arise directly from a difference in the ratio of bandwidth to  correlation strength. 

Further insight is obtained from the bare  hybridization function $\mathbf{\Delta}^0$.  This is the crucial input to the single-site and cluster dynamical mean-field calculations. It is defined in terms of the bare  local Green function ${\mathbf G}^0_{loc}$ defined as the matrix-elements of the band theory  Green function $\mathbf G_{band}=\left[(\omega+\mu){\mathbf 1}-\mathbf H_{Kohn-Sham}\right]^{-1}$ onto the maximially localized Wannier functions associated with the  two $J_\mathrm{eff}=1/2$ states of  single Ir site  (sDMFT) or the 8 states associated with a tetrahedron (CDMFT). Then
\begin{equation}
{\mathbf \Delta}^0(\omega)=(\omega+\mu){\mathbf 1}-\left({\mathbf G}^0_{loc}(\omega)\right)^{-1}
\label{Deltadef}
\end{equation}

\begin{figure}
\begin{center}
\includegraphics[width=0.9\columnwidth]{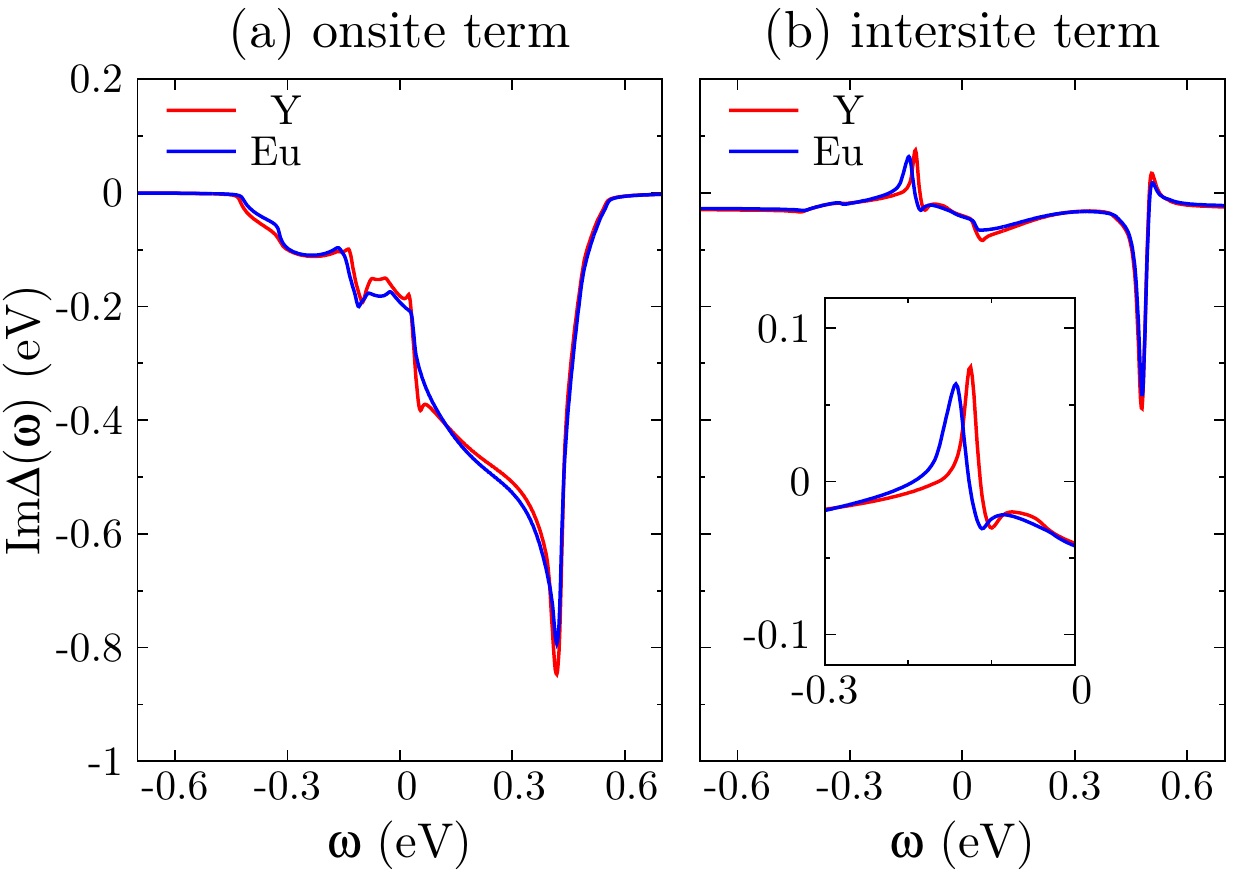}
\caption{On-site (panel (a)) and intersite (panel (b)) contributions to the imaginary part of the bare hybridization function $\mathbf{\Delta}(\omega)$ (defined in the main text). The two panels share the same ordinate. The inset of panel (b) shows an expanded view of the range $-0.3\lesssim \omega \lesssim 0$ highlighting the shift of the peak.}
\label{hyb}
\end{center}
\end{figure}

In the single-site DMFT approximation $\mathbf{\Delta}^0$ is a $2\times 2$ matrix; in the paramagnetic state time reversal invariance ensures that it is proportional to the unit matrix, so is described by one function of frequency, whose imaginary part is shown in  Fig. \ref{hyb}(a) for the Y and Eu materials. We see that the single-site DMFT bare hybridization functions for the two compounds  are essentially identical, explaining why the single-site DMFT approximation predicts the same properties for the two compounds.

\begin{figure}
\begin{center}
\includegraphics[width=0.9\columnwidth]{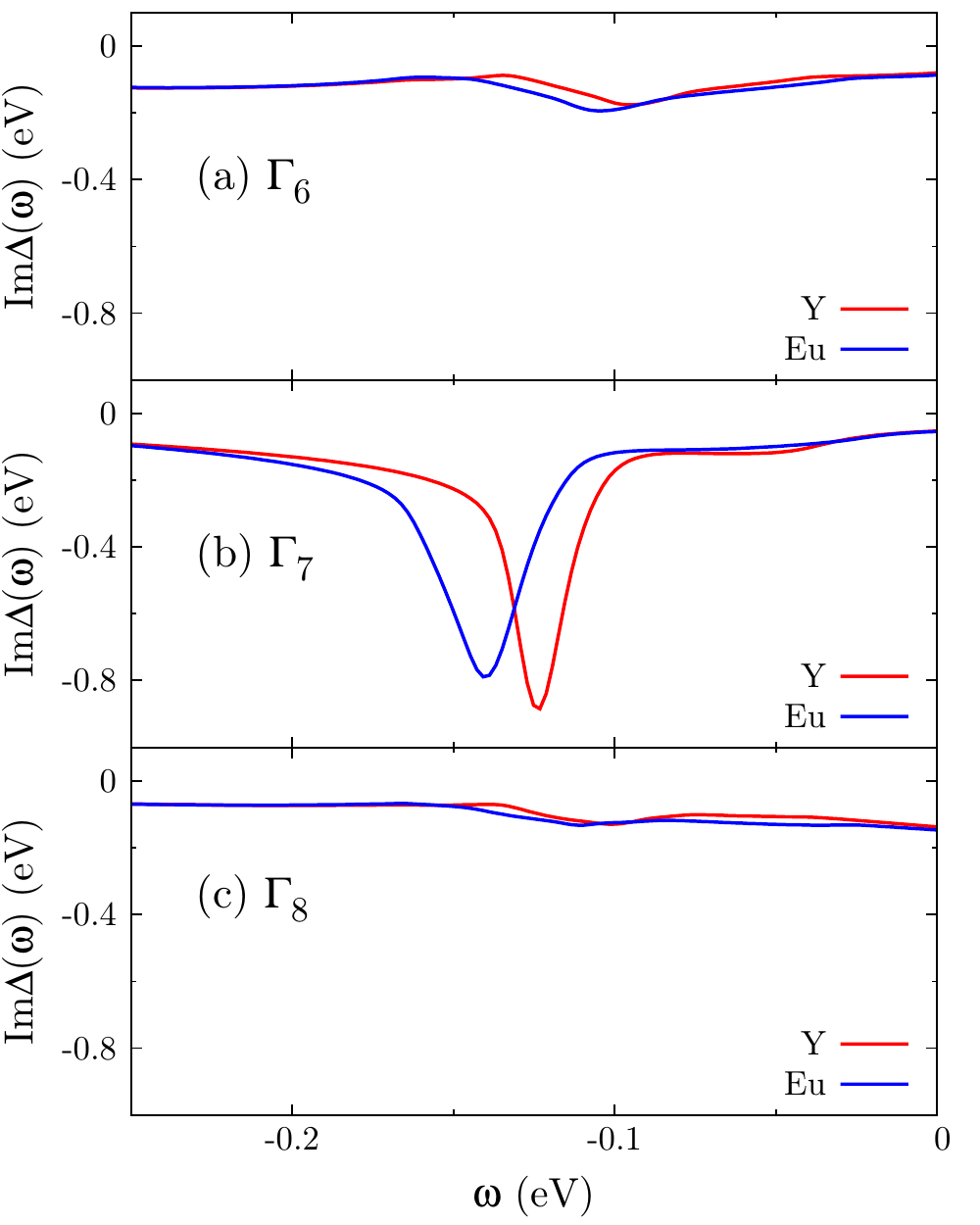}
\caption{Projection of the bare hybridization function $\mathbf{\Delta}(\omega)$ onto the irreducible representations $\Gamma_6$, $\Gamma_7$ and $\Gamma_8$ of the tetrahedral point group. The three panels share the same horizontal axis. Panel (b) shows that the material-dependent peak appears only in the $\Gamma_7$ representation.}
\label{hyb-rotated}
\end{center}
\end{figure}

We now turn to the CDMFT case, where $\mathbf{\Delta}^0$ is an $8\times 8$ matrix. It is useful to decompose this matrix into irreducible representations of the double group describing the point symmetries of the pyrochlore iridate structure (which is a 4-site Ir tetrahedron in our impurity model) in the paramagnetic phase. In the notation of Ref. ~\cite{book}, the relevant representations are $\Gamma_6$ and $\Gamma_7$ (doublet) and $\Gamma_8$ (quartet) and we have
\begin{equation}
\mathbf{\Delta}^0(\omega)=\sum_{a=\Gamma_{6,7,8}}\Delta_a^0(\omega)\sum_{\lambda_a}\left|\psi^a_{\lambda_a}\right\rangle\left\langle\psi^a_{\lambda_a}\right|
\end{equation}
Fig. \ref{hyb-rotated} shows the imaginary part of the projections of the hybridization function onto the three irreducible representations. We see that the significant material difference appears primarily in the $\Gamma_7$ representation.

\begin{figure}[t]
\begin{center}
\includegraphics[width=0.9\columnwidth]{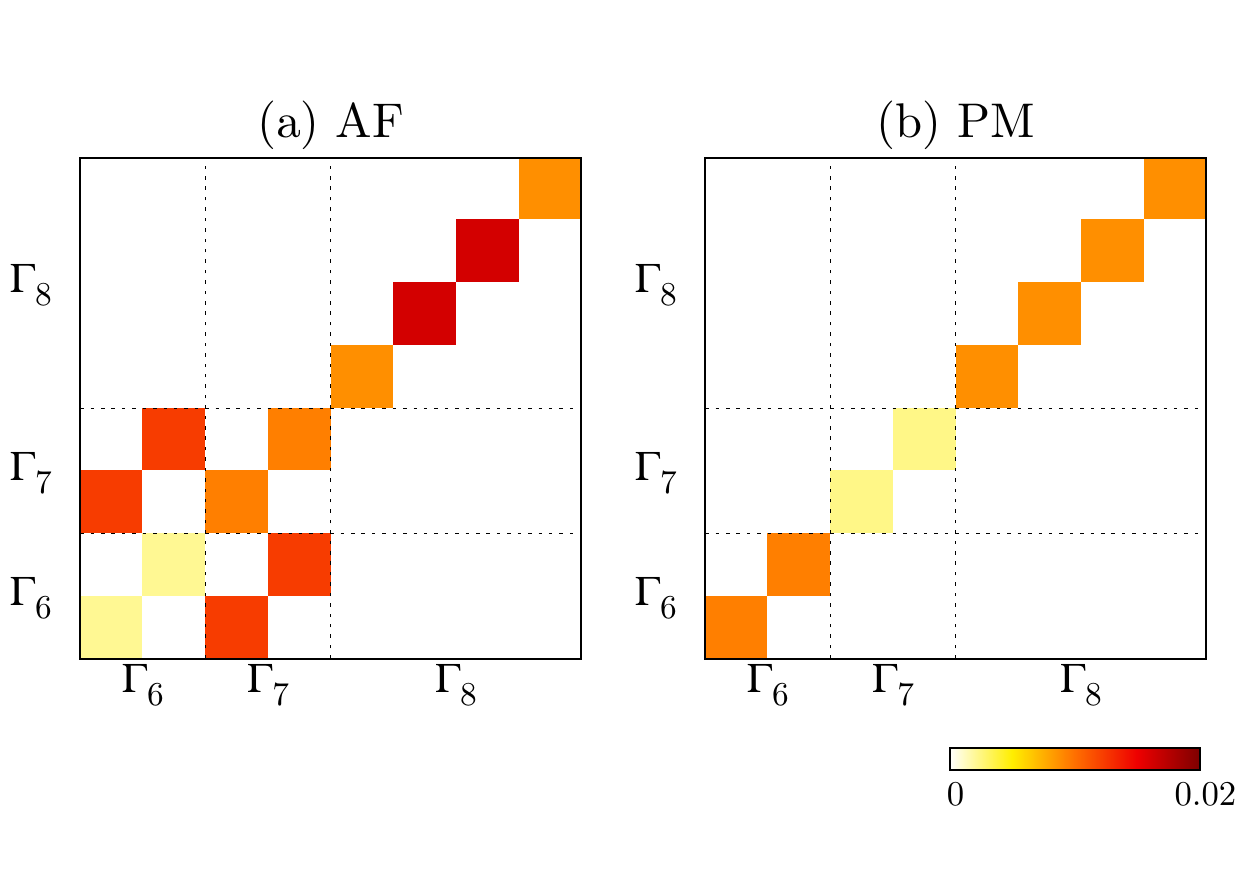}
\caption{Material difference of the hybridization function for Y-227 and Eu-227 in the case of $U$=1.1eV and $\omega_0$=0.001eV in (a) antiferromagnetic state and (b) paramagnetic state. The color for a given matrix element in a given solution is defined as $||\Delta^Y_{ij}(i\omega_0)|-|\Delta^{Eu}_{ij}(i\omega_0)||$.}
\label{AF-PM-hybrid}
\end{center}
\end{figure}

The small material-dependence of the bare hybridization function is amplified by the many-body physics, leading to the material dependence found in the calculations. This may be seen in results for the interaction hybridization function, defined by using the full interacting DFT+DMFT Green function in Eq.~(\ref{Deltadef}).  In Fig. \ref{AF-PM-hybrid}, we plot the absolute value of the magnitude difference between Y-227 and Eu-227 in the hybridization function at the lowest Matsubara frequency $\omega_0$ for every matrix element in the case of antiferromagnetic state and paramagnetic state respectively. The time-reversal symmetry breaking associated with the antiferromagnetic state changes the structure of the hybridization function,  among other things leading  to mixing between the $\Gamma_6$ and $\Gamma_7$. This mixing, as well as a change in the structure of the diagonal elements of the  $\Gamma_8$ representation,  exhibits  particularly strong material dependence.

\section{Conclusion} \label{conclusion}
This work presents  density functional plus dynamical mean-field calculations of pyrochlore iridates  based on a representation in which the important interactions occurred between electrons in the frontier ($J_\mathrm{eff}=1/2$) orbitals. The downfolding to the $J_\mathrm{eff}=1/2$ manifold is an approximation in this work, which is justified by the gap separating the $J_\mathrm{eff}=1/2$ and $J_\mathrm{eff}=3/2$ manifolds. Nevertheless, the investigation of the effects of the $J_\mathrm{eff}=3/2$ manifold \cite{lda-dmft-Y,lda-dmft-R-dep} is an important issue for future research. Important features of our work are the incorporation of spin-orbit coupling and the use of cluster dynamical mean-field methods.  An important parameter in the dynamical mean-field calculations is the effective interaction $U$  amongst electrons in the correlated orbitals. Its value is strongly affected by screening and by the choice of correlated orbitals; at present it should be determined phenomenologically.  We investigated the qualitative behavior as $U$ is varied. As $U$ is increased from zero our cluster dynamical mean-field calculations reveal a first-order magnetic transition from a paramagnetic metal to antiferromagnetic  metal with AIAO order and Weyl nodes followed by  a transition to a topologically trivial insulator with  AIAO order. The prediction of the first-order nature of magnetic transition could be tested by measurements of the pressure dependence of the staggered magnetization or of the ordered moment.  This sequence of phases was previously reported in Hartree-Fock \cite{tb-HF1, tb-HF2} calculations. DFT+U calculations \cite{dft} found these phases and also suggested the possibility of an axion insulator phase, which was found in model system CDMFT \cite{AI} calculations, but not here. Ref. ~\cite{dft} also explained the association between the metal-insulator transition and the change in topology. Calculations based on the single-site DMFT approximation \cite{lda-dmft-Y, lda-dmft-R-dep} reported a direct transition from paramagnetic metal to topologically trivial antiferromagnetic insulator, with the Weyl metal phase being absent.  

We considered mapping the calculated results onto experiment in two ways. First, we consider the magnitude of the antiferromagnetic and paramagnetic phase gaps, as determined from the resistivity and optics.  We found that the single-site dynamical mean-field theory could not account for the small (but non-vanishing) gaps inferred from paramagnetic phase resistivity measurements on many pyrochlore iridates, whereas the cluster dynamical mean-field calculations could. These considerations suggest that the materials are reasonably well described by DFT+CDMFT calculations with a frontier-orbital $U\approx 1$eV. An unresolved difficulty with this point of view is that there is little experimental support for the relatively wide Weyl metal regime predicted by our CDMFT results. Instead, experiment seems to favor a picture more similar to the single-site phase diagram \cite{lda-dmft-Y, lda-dmft-R-dep}, with a direct transition from paramagnetic metal to topologically trivial insulator, although a few experiments report indications of Weyl semimetal phases \cite{Ueda16, WSM-optical-Eu}. Further experimental study of this issue would be desirable.

We also investigated  the dependence of material properties on the choice of rare earth ion or Y. Some aspects of the behavior are consistent with the conventional understanding developed in the context of the ABO$_3$ perovskite family of materials.  The strongly insulating behavior of Lu$_2$Ir$_2$O$_7$ is found to be associated with a narrower bandwidth arising from the octahedral distortion driven by the smaller size of the Lu and is well described by both the single-site and cluster DMFT calculations. However, the pronounced difference in properties between Y$_2$Ir$_2$O$_7$ and Eu$_2$Ir$_2$O$_7$ seems not to be related to a bandwidth effect. Indeed,  the paramagnetic-phase single-site  DMFT approximation predicts almost identical gaps for the materials, because the bare hybridization functions are almost identical, and it is only when antiferromagnetism is considered that a significant material difference appears.  The noticeable differences in paramagnetic phase resistivities of the Y and Eu materials then suggests that antiferromagnetic fluctuations play an important role in determining the physical properties in the paramagnetic phase. This idea is further supported by optical data indicating that raising the temperature above the N\'eel temperature leads to a large broadening of the gap but only a modest decrease. A wide fluctuation regime is expected in low dimensional materials, but is remarkable in three dimensional materials such as the pyrochlore iridates. 

The results have interesting implications for the dynamical mean-field method. One basic implication is that the single-site approximation may not be entirely adequate to describe the paramagnetic phase physics of the pyrochlore iridates. Cluster DMFT calculations include intersite correlations and procude the requisite small gaps and may therefore provide a more natural description of the physics. Material differences appear in  terms of the bare hybridization function associated with intersite correlations.  However, our paramagnetic-phase CDMFT calculations, while producing the requisite small gap, still underestimate the Y-Eu material difference. We believe that this failure is a limitation of the exact diagonalization method that we have used to solve the dynamical mean-field equations. In the implementation used here, ED is a ground state method. Since in the relevant parameter range the actual ground state is antiferromagnetic, an averaging procedure must be employed to force the hybridization function to have the symmetries required in the paramagnetic state. We believe that this suppresses the magnetic fluctuations. A continuous time quantum Monte Carlo calculation that can explicitly address the high-$T$ phase would be worth performing. However, even at $T$=0, the factor of two difference in reported gap values could only be explained via a 10\% change of $U$.

\section*{Acknowledgement} 
We thank Chris A. Marianetti, Heung-Sik Kim and Hongbin Zhang for helpful discussions. This research was supported by the Basic Energy Sciences Division of the DOE Office of Science under grant ER-046169. The computing resources were provided by the Laboratory Computing Resource Center at Argonne National Laboratory, the Extreme Science and Engineering Discovery Environment (XSEDE) via the Grant No. ACI-1053575 and the National Energy Research Scientific Computing Center, a DOE Office of Science User Facility supported by the Office of Science of the U.S. Department of Energy under Contract No. DE-AC02-05CH11231.

\section*{Appendix: Dependence on the number of bath orbitals \label{bath-orbitals}}
\begin{figure}
\begin{center}
\includegraphics[width=0.9\columnwidth]{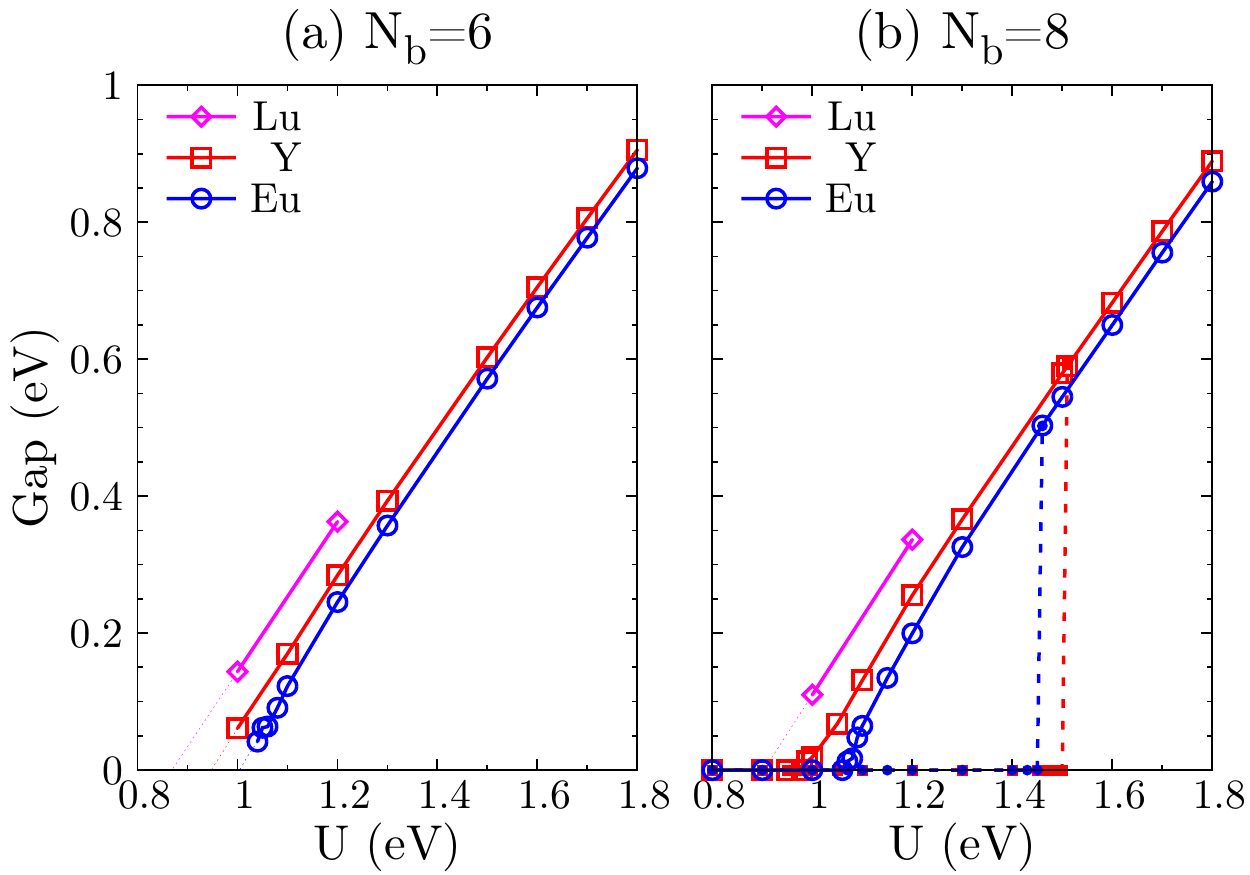}
\caption{Gap size as a function of the interaction strength with (a) $N_b=6$ and (b) $N_b=8$. The two panels share the same ordinate.}
\label{AF-nb}
\end{center}
\end{figure}
In this appendix, we discuss the dependence of the spectral gap on the number of bath orbitals ($N_b$) within CDMFT calculations. Unless there is infinite number of bath orbitals, how well the fitting of the Gaussian Weiss field is depends on $N_b$, and therefore the results exhibits a dependence on $N_b$ and will converge when $N_b$ is large enough. In our 4-site cluster calculations, there is an upper limit for $N_b$ ($N_b\leqslant8$), therefore, the convergence of $N_b$ is not able to be verified. However, we can get some trends from the comparison of the spectral gap obtain with $N_b=6$ and $N_b=8$, even though $N_b=6$ is not considered that good given that there are 4 correlated orbitals in our calculations. As presented by Fig. \ref{AF-nb}, for a given compound, the spectral gap reduces as $N_b$ increases and the error due to $N_b$ also reduces as $U$ increases. Furthermore, it turns out that the material difference obtained with $N_b=8$ is slightly larger than that with $N_b=6$ for any $U$, especially near the transition.

\bibliography{Refs}

\begin{thebibliography}{50}%
\makeatletter
\providecommand \@ifxundefined [1]{%
 \@ifx{#1\undefined}
}%
\providecommand \@ifnum [1]{%
 \ifnum #1\expandafter \@firstoftwo
 \else \expandafter \@secondoftwo
 \fi
}%
\providecommand \@ifx [1]{%
 \ifx #1\expandafter \@firstoftwo
 \else \expandafter \@secondoftwo
 \fi
}%
\providecommand \natexlab [1]{#1}%
\providecommand \enquote  [1]{``#1''}%
\providecommand \bibnamefont  [1]{#1}%
\providecommand \bibfnamefont [1]{#1}%
\providecommand \citenamefont [1]{#1}%
\providecommand \href@noop [0]{\@secondoftwo}%
\providecommand \href [0]{\begingroup \@sanitize@url \@href}%
\providecommand \@href[1]{\@@startlink{#1}\@@href}%
\providecommand \@@href[1]{\endgroup#1\@@endlink}%
\providecommand \@sanitize@url [0]{\catcode `\\12\catcode `\$12\catcode
  `\&12\catcode `\#12\catcode `\^12\catcode `\_12\catcode `\%12\relax}%
\providecommand \@@startlink[1]{}%
\providecommand \@@endlink[0]{}%
\providecommand \url  [0]{\begingroup\@sanitize@url \@url }%
\providecommand \@url [1]{\endgroup\@href {#1}{\urlprefix }}%
\providecommand \urlprefix  [0]{URL }%
\providecommand \Eprint [0]{\href }%
\providecommand \doibase [0]{http://dx.doi.org/}%
\providecommand \selectlanguage [0]{\@gobble}%
\providecommand \bibinfo  [0]{\@secondoftwo}%
\providecommand \bibfield  [0]{\@secondoftwo}%
\providecommand \translation [1]{[#1]}%
\providecommand \BibitemOpen [0]{}%
\providecommand \bibitemStop [0]{}%
\providecommand \bibitemNoStop [0]{.\EOS\space}%
\providecommand \EOS [0]{\spacefactor3000\relax}%
\providecommand \BibitemShut  [1]{\csname bibitem#1\endcsname}%
\let\auto@bib@innerbib\@empty
\bibitem [{\citenamefont {Pesin}\ and\ \citenamefont
  {Balents}(2010)}]{Pesin2010}%
  \BibitemOpen
  \bibfield  {author} {\bibinfo {author} {\bibfnamefont {D.}~\bibnamefont
  {Pesin}}\ and\ \bibinfo {author} {\bibfnamefont {L.}~\bibnamefont
  {Balents}},\ }\href {\doibase 10.1038/nphys1606} {\bibfield  {journal}
  {\bibinfo  {journal} {Nat Phys}\ }\textbf {\bibinfo {volume} {6}},\ \bibinfo
  {pages} {376} (\bibinfo {year} {2010})}\BibitemShut {NoStop}%
\bibitem [{\citenamefont {Nakatsuji}\ \emph {et~al.}(2006)\citenamefont
  {Nakatsuji}, \citenamefont {Machida}, \citenamefont {Maeno}, \citenamefont
  {Tayama}, \citenamefont {Sakakibara}, \citenamefont {Duijn}, \citenamefont
  {Balicas}, \citenamefont {Millican}, \citenamefont {Macaluso},\ and\
  \citenamefont {Chan}}]{spinliqiud-prl}%
  \BibitemOpen
  \bibfield  {author} {\bibinfo {author} {\bibfnamefont {S.}~\bibnamefont
  {Nakatsuji}}, \bibinfo {author} {\bibfnamefont {Y.}~\bibnamefont {Machida}},
  \bibinfo {author} {\bibfnamefont {Y.}~\bibnamefont {Maeno}}, \bibinfo
  {author} {\bibfnamefont {T.}~\bibnamefont {Tayama}}, \bibinfo {author}
  {\bibfnamefont {T.}~\bibnamefont {Sakakibara}}, \bibinfo {author}
  {\bibfnamefont {J.~v.}\ \bibnamefont {Duijn}}, \bibinfo {author}
  {\bibfnamefont {L.}~\bibnamefont {Balicas}}, \bibinfo {author} {\bibfnamefont
  {J.~N.}\ \bibnamefont {Millican}}, \bibinfo {author} {\bibfnamefont {R.~T.}\
  \bibnamefont {Macaluso}}, \ and\ \bibinfo {author} {\bibfnamefont {J.~Y.}\
  \bibnamefont {Chan}},\ }\href {\doibase 10.1103/PhysRevLett.96.087204}
  {\bibfield  {journal} {\bibinfo  {journal} {Phys. Rev. Lett.}\ }\textbf
  {\bibinfo {volume} {96}},\ \bibinfo {pages} {087204} (\bibinfo {year}
  {2006})}\BibitemShut {NoStop}%
\bibitem [{\citenamefont {Machida}\ \emph {et~al.}(2010)\citenamefont
  {Machida}, \citenamefont {Nakatsuji}, \citenamefont {Onoda}, \citenamefont
  {Tayama},\ and\ \citenamefont {Sakakibara}}]{spinliqiud-nature}%
  \BibitemOpen
  \bibfield  {author} {\bibinfo {author} {\bibfnamefont {Y.}~\bibnamefont
  {Machida}}, \bibinfo {author} {\bibfnamefont {S.}~\bibnamefont {Nakatsuji}},
  \bibinfo {author} {\bibfnamefont {S.}~\bibnamefont {Onoda}}, \bibinfo
  {author} {\bibfnamefont {T.}~\bibnamefont {Tayama}}, \ and\ \bibinfo {author}
  {\bibfnamefont {T.}~\bibnamefont {Sakakibara}},\ }\href {\doibase
  10.1038/nature08680} {\bibfield  {journal} {\bibinfo  {journal} {Nature}\
  }\textbf {\bibinfo {volume} {463}},\ \bibinfo {pages} {210} (\bibinfo {year}
  {2010})}\BibitemShut {NoStop}%
\bibitem [{\citenamefont {Wan}\ \emph {et~al.}(2011)\citenamefont {Wan},
  \citenamefont {Turner}, \citenamefont {Vishwanath},\ and\ \citenamefont
  {Savrasov}}]{dft}%
  \BibitemOpen
  \bibfield  {author} {\bibinfo {author} {\bibfnamefont {X.}~\bibnamefont
  {Wan}}, \bibinfo {author} {\bibfnamefont {A.~M.}\ \bibnamefont {Turner}},
  \bibinfo {author} {\bibfnamefont {A.}~\bibnamefont {Vishwanath}}, \ and\
  \bibinfo {author} {\bibfnamefont {S.~Y.}\ \bibnamefont {Savrasov}},\ }\href
  {\doibase 10.1103/PhysRevB.83.205101} {\bibfield  {journal} {\bibinfo
  {journal} {Phys. Rev. B}\ }\textbf {\bibinfo {volume} {83}},\ \bibinfo
  {pages} {205101} (\bibinfo {year} {2011})}\BibitemShut {NoStop}%
\bibitem [{\citenamefont {Witczak-Krempa}\ and\ \citenamefont
  {Kim}(2012)}]{tb-HF1}%
  \BibitemOpen
  \bibfield  {author} {\bibinfo {author} {\bibfnamefont {W.}~\bibnamefont
  {Witczak-Krempa}}\ and\ \bibinfo {author} {\bibfnamefont {Y.~B.}\
  \bibnamefont {Kim}},\ }\href {\doibase 10.1103/PhysRevB.85.045124} {\bibfield
   {journal} {\bibinfo  {journal} {Phys. Rev. B}\ }\textbf {\bibinfo {volume}
  {85}},\ \bibinfo {pages} {045124} (\bibinfo {year} {2012})}\BibitemShut
  {NoStop}%
\bibitem [{\citenamefont {Witczak-Krempa}\ \emph {et~al.}(2013)\citenamefont
  {Witczak-Krempa}, \citenamefont {Go},\ and\ \citenamefont {Kim}}]{tb-HF2}%
  \BibitemOpen
  \bibfield  {author} {\bibinfo {author} {\bibfnamefont {W.}~\bibnamefont
  {Witczak-Krempa}}, \bibinfo {author} {\bibfnamefont {A.}~\bibnamefont {Go}},
  \ and\ \bibinfo {author} {\bibfnamefont {Y.~B.}\ \bibnamefont {Kim}},\ }\href
  {\doibase 10.1103/PhysRevB.87.155101} {\bibfield  {journal} {\bibinfo
  {journal} {Phys. Rev. B}\ }\textbf {\bibinfo {volume} {87}},\ \bibinfo
  {pages} {155101} (\bibinfo {year} {2013})}\BibitemShut {NoStop}%
\bibitem [{\citenamefont {Go}\ \emph {et~al.}(2012)\citenamefont {Go},
  \citenamefont {Witczak-Krempa}, \citenamefont {Jeon}, \citenamefont {Park},\
  and\ \citenamefont {Kim}}]{AI}%
  \BibitemOpen
  \bibfield  {author} {\bibinfo {author} {\bibfnamefont {A.}~\bibnamefont
  {Go}}, \bibinfo {author} {\bibfnamefont {W.}~\bibnamefont {Witczak-Krempa}},
  \bibinfo {author} {\bibfnamefont {G.~S.}\ \bibnamefont {Jeon}}, \bibinfo
  {author} {\bibfnamefont {K.}~\bibnamefont {Park}}, \ and\ \bibinfo {author}
  {\bibfnamefont {Y.~B.}\ \bibnamefont {Kim}},\ }\href {\doibase
  10.1103/PhysRevLett.109.066401} {\bibfield  {journal} {\bibinfo  {journal}
  {Phys. Rev. Lett.}\ }\textbf {\bibinfo {volume} {109}},\ \bibinfo {pages}
  {066401} (\bibinfo {year} {2012})}\BibitemShut {NoStop}%
\bibitem [{\citenamefont {Yanagishima}\ and\ \citenamefont
  {Maeno}(2001)}]{metal-nonmetal-Pr-Nd-Sm-Eu}%
  \BibitemOpen
  \bibfield  {author} {\bibinfo {author} {\bibfnamefont {D.}~\bibnamefont
  {Yanagishima}}\ and\ \bibinfo {author} {\bibfnamefont {Y.}~\bibnamefont
  {Maeno}},\ }\href {\doibase 10.1143/JPSJ.70.2880} {\bibfield  {journal}
  {\bibinfo  {journal} {Journal of the Physical Society of Japan}\ }\textbf
  {\bibinfo {volume} {70}},\ \bibinfo {pages} {2880} (\bibinfo {year}
  {2001})}\BibitemShut {NoStop}%
\bibitem [{\citenamefont {Matsuhira}\ \emph {et~al.}(2007)\citenamefont
  {Matsuhira}, \citenamefont {Wakeshima}, \citenamefont {Nakanishi},
  \citenamefont {Yamada}, \citenamefont {Nakamura}, \citenamefont {Kawano},
  \citenamefont {Takagi},\ and\ \citenamefont {Hinatsu}}]{MIT-Nd-Sm-Eu}%
  \BibitemOpen
  \bibfield  {author} {\bibinfo {author} {\bibfnamefont {K.}~\bibnamefont
  {Matsuhira}}, \bibinfo {author} {\bibfnamefont {M.}~\bibnamefont
  {Wakeshima}}, \bibinfo {author} {\bibfnamefont {R.}~\bibnamefont
  {Nakanishi}}, \bibinfo {author} {\bibfnamefont {T.}~\bibnamefont {Yamada}},
  \bibinfo {author} {\bibfnamefont {A.}~\bibnamefont {Nakamura}}, \bibinfo
  {author} {\bibfnamefont {W.}~\bibnamefont {Kawano}}, \bibinfo {author}
  {\bibfnamefont {S.}~\bibnamefont {Takagi}}, \ and\ \bibinfo {author}
  {\bibfnamefont {Y.}~\bibnamefont {Hinatsu}},\ }\href {\doibase
  10.1143/JPSJ.76.043706} {\bibfield  {journal} {\bibinfo  {journal} {Journal
  of the Physical Society of Japan}\ }\textbf {\bibinfo {volume} {76}},\
  \bibinfo {pages} {043706} (\bibinfo {year} {2007})}\BibitemShut {NoStop}%
\bibitem [{\citenamefont {Matsuhira}\ \emph {et~al.}(2011)\citenamefont
  {Matsuhira}, \citenamefont {Wakeshima}, \citenamefont {Hinatsu},\ and\
  \citenamefont {Takagi}}]{MIT-Nd-Sm-Eu-Gd-Tb-Dy-Ho}%
  \BibitemOpen
  \bibfield  {author} {\bibinfo {author} {\bibfnamefont {K.}~\bibnamefont
  {Matsuhira}}, \bibinfo {author} {\bibfnamefont {M.}~\bibnamefont
  {Wakeshima}}, \bibinfo {author} {\bibfnamefont {Y.}~\bibnamefont {Hinatsu}},
  \ and\ \bibinfo {author} {\bibfnamefont {S.}~\bibnamefont {Takagi}},\ }\href
  {\doibase 10.1143/JPSJ.80.094701} {\bibfield  {journal} {\bibinfo  {journal}
  {Journal of the Physical Society of Japan}\ }\textbf {\bibinfo {volume}
  {80}},\ \bibinfo {pages} {094701} (\bibinfo {year} {2011})}\BibitemShut
  {NoStop}%
\bibitem [{\citenamefont {Zhao}\ \emph {et~al.}(2011)\citenamefont {Zhao},
  \citenamefont {Mackie}, \citenamefont {MacLaughlin}, \citenamefont {Bernal},
  \citenamefont {Ishikawa}, \citenamefont {Ohta},\ and\ \citenamefont
  {Nakatsuji}}]{MIT-mag-Eu1}%
  \BibitemOpen
  \bibfield  {author} {\bibinfo {author} {\bibfnamefont {S.}~\bibnamefont
  {Zhao}}, \bibinfo {author} {\bibfnamefont {J.~M.}\ \bibnamefont {Mackie}},
  \bibinfo {author} {\bibfnamefont {D.~E.}\ \bibnamefont {MacLaughlin}},
  \bibinfo {author} {\bibfnamefont {O.~O.}\ \bibnamefont {Bernal}}, \bibinfo
  {author} {\bibfnamefont {J.~J.}\ \bibnamefont {Ishikawa}}, \bibinfo {author}
  {\bibfnamefont {Y.}~\bibnamefont {Ohta}}, \ and\ \bibinfo {author}
  {\bibfnamefont {S.}~\bibnamefont {Nakatsuji}},\ }\href {\doibase
  10.1103/PhysRevB.83.180402} {\bibfield  {journal} {\bibinfo  {journal} {Phys.
  Rev. B}\ }\textbf {\bibinfo {volume} {83}},\ \bibinfo {pages} {180402}
  (\bibinfo {year} {2011})}\BibitemShut {NoStop}%
\bibitem [{\citenamefont {Ishikawa}\ \emph {et~al.}(2012)\citenamefont
  {Ishikawa}, \citenamefont {O'Farrell},\ and\ \citenamefont
  {Nakatsuji}}]{MIT-mag-Eu2}%
  \BibitemOpen
  \bibfield  {author} {\bibinfo {author} {\bibfnamefont {J.~J.}\ \bibnamefont
  {Ishikawa}}, \bibinfo {author} {\bibfnamefont {E.~C.~T.}\ \bibnamefont
  {O'Farrell}}, \ and\ \bibinfo {author} {\bibfnamefont {S.}~\bibnamefont
  {Nakatsuji}},\ }\href {\doibase 10.1103/PhysRevB.85.245109} {\bibfield
  {journal} {\bibinfo  {journal} {Phys. Rev. B}\ }\textbf {\bibinfo {volume}
  {85}},\ \bibinfo {pages} {245109} (\bibinfo {year} {2012})}\BibitemShut
  {NoStop}%
\bibitem [{\citenamefont {Disseler}\ \emph
  {et~al.}(2012{\natexlab{a}})\citenamefont {Disseler}, \citenamefont {Dhital},
  \citenamefont {Amato}, \citenamefont {Giblin}, \citenamefont {de~la Cruz},
  \citenamefont {Wilson},\ and\ \citenamefont {Graf}}]{MIT-mag-Yb}%
  \BibitemOpen
  \bibfield  {author} {\bibinfo {author} {\bibfnamefont {S.~M.}\ \bibnamefont
  {Disseler}}, \bibinfo {author} {\bibfnamefont {C.}~\bibnamefont {Dhital}},
  \bibinfo {author} {\bibfnamefont {A.}~\bibnamefont {Amato}}, \bibinfo
  {author} {\bibfnamefont {S.~R.}\ \bibnamefont {Giblin}}, \bibinfo {author}
  {\bibfnamefont {C.}~\bibnamefont {de~la Cruz}}, \bibinfo {author}
  {\bibfnamefont {S.~D.}\ \bibnamefont {Wilson}}, \ and\ \bibinfo {author}
  {\bibfnamefont {M.~J.}\ \bibnamefont {Graf}},\ }\href {\doibase
  10.1103/PhysRevB.86.014428} {\bibfield  {journal} {\bibinfo  {journal} {Phys.
  Rev. B}\ }\textbf {\bibinfo {volume} {86}},\ \bibinfo {pages} {014428}
  (\bibinfo {year} {2012}{\natexlab{a}})}\BibitemShut {NoStop}%
\bibitem [{\citenamefont {Guo}\ \emph {et~al.}(2013)\citenamefont {Guo},
  \citenamefont {Matsuhira}, \citenamefont {Kawasaki}, \citenamefont
  {Wakeshima}, \citenamefont {Hinatsu}, \citenamefont {Watanabe},\ and\
  \citenamefont {Xu}}]{MIT-mag-Nd}%
  \BibitemOpen
  \bibfield  {author} {\bibinfo {author} {\bibfnamefont {H.}~\bibnamefont
  {Guo}}, \bibinfo {author} {\bibfnamefont {K.}~\bibnamefont {Matsuhira}},
  \bibinfo {author} {\bibfnamefont {I.}~\bibnamefont {Kawasaki}}, \bibinfo
  {author} {\bibfnamefont {M.}~\bibnamefont {Wakeshima}}, \bibinfo {author}
  {\bibfnamefont {Y.}~\bibnamefont {Hinatsu}}, \bibinfo {author} {\bibfnamefont
  {I.}~\bibnamefont {Watanabe}}, \ and\ \bibinfo {author} {\bibfnamefont
  {Z.-a.}\ \bibnamefont {Xu}},\ }\href {\doibase 10.1103/PhysRevB.88.060411}
  {\bibfield  {journal} {\bibinfo  {journal} {Phys. Rev. B}\ }\textbf {\bibinfo
  {volume} {88}},\ \bibinfo {pages} {060411} (\bibinfo {year}
  {2013})}\BibitemShut {NoStop}%
\bibitem [{\citenamefont {Ueda}\ \emph {et~al.}(2016)\citenamefont {Ueda},
  \citenamefont {Fujioka},\ and\ \citenamefont {Tokura}}]{Ueda16}%
  \BibitemOpen
  \bibfield  {author} {\bibinfo {author} {\bibfnamefont {K.}~\bibnamefont
  {Ueda}}, \bibinfo {author} {\bibfnamefont {J.}~\bibnamefont {Fujioka}}, \
  and\ \bibinfo {author} {\bibfnamefont {Y.}~\bibnamefont {Tokura}},\ }\href
  {\doibase 10.1103/PhysRevB.93.245120} {\bibfield  {journal} {\bibinfo
  {journal} {Phys. Rev. B}\ }\textbf {\bibinfo {volume} {93}},\ \bibinfo
  {pages} {245120} (\bibinfo {year} {2016})}\BibitemShut {NoStop}%
\bibitem [{\citenamefont {Taira}\ \emph {et~al.}(2001)\citenamefont {Taira},
  \citenamefont {Wakeshima},\ and\ \citenamefont {Hinatsu}}]{mag-Y-Sm-Eu-Lu}%
  \BibitemOpen
  \bibfield  {author} {\bibinfo {author} {\bibfnamefont {N.}~\bibnamefont
  {Taira}}, \bibinfo {author} {\bibfnamefont {M.}~\bibnamefont {Wakeshima}}, \
  and\ \bibinfo {author} {\bibfnamefont {Y.}~\bibnamefont {Hinatsu}},\ }\href
  {http://stacks.iop.org/0953-8984/13/i=23/a=312} {\bibfield  {journal}
  {\bibinfo  {journal} {Journal of Physics: Condensed Matter}\ }\textbf
  {\bibinfo {volume} {13}},\ \bibinfo {pages} {5527} (\bibinfo {year}
  {2001})}\BibitemShut {NoStop}%
\bibitem [{\citenamefont {Kumar}\ and\ \citenamefont
  {Pramanik}(2016)}]{glass-like}%
  \BibitemOpen
  \bibfield  {author} {\bibinfo {author} {\bibfnamefont {H.}~\bibnamefont
  {Kumar}}\ and\ \bibinfo {author} {\bibfnamefont {A.}~\bibnamefont
  {Pramanik}},\ }\href {\doibase http://dx.doi.org/10.1016/j.jmmm.2016.02.033}
  {\bibfield  {journal} {\bibinfo  {journal} {Journal of Magnetism and Magnetic
  Materials}\ }\textbf {\bibinfo {volume} {409}},\ \bibinfo {pages} {20}
  (\bibinfo {year} {2016})}\BibitemShut {NoStop}%
\bibitem [{\citenamefont {Disseler}\ \emph
  {et~al.}(2012{\natexlab{b}})\citenamefont {Disseler}, \citenamefont {Dhital},
  \citenamefont {Hogan}, \citenamefont {Amato}, \citenamefont {Giblin},
  \citenamefont {de~la Cruz}, \citenamefont {Daoud-Aladine}, \citenamefont
  {Wilson},\ and\ \citenamefont {Graf}}]{neutron-Nd}%
  \BibitemOpen
  \bibfield  {author} {\bibinfo {author} {\bibfnamefont {S.~M.}\ \bibnamefont
  {Disseler}}, \bibinfo {author} {\bibfnamefont {C.}~\bibnamefont {Dhital}},
  \bibinfo {author} {\bibfnamefont {T.~C.}\ \bibnamefont {Hogan}}, \bibinfo
  {author} {\bibfnamefont {A.}~\bibnamefont {Amato}}, \bibinfo {author}
  {\bibfnamefont {S.~R.}\ \bibnamefont {Giblin}}, \bibinfo {author}
  {\bibfnamefont {C.}~\bibnamefont {de~la Cruz}}, \bibinfo {author}
  {\bibfnamefont {A.}~\bibnamefont {Daoud-Aladine}}, \bibinfo {author}
  {\bibfnamefont {S.~D.}\ \bibnamefont {Wilson}}, \ and\ \bibinfo {author}
  {\bibfnamefont {M.~J.}\ \bibnamefont {Graf}},\ }\href {\doibase
  10.1103/PhysRevB.85.174441} {\bibfield  {journal} {\bibinfo  {journal} {Phys.
  Rev. B}\ }\textbf {\bibinfo {volume} {85}},\ \bibinfo {pages} {174441}
  (\bibinfo {year} {2012}{\natexlab{b}})}\BibitemShut {NoStop}%
\bibitem [{\citenamefont {Shapiro}\ \emph {et~al.}(2012)\citenamefont
  {Shapiro}, \citenamefont {Riggs}, \citenamefont {Stone}, \citenamefont {de~la
  Cruz}, \citenamefont {Chi}, \citenamefont {Podlesnyak},\ and\ \citenamefont
  {Fisher}}]{neutron-Y}%
  \BibitemOpen
  \bibfield  {author} {\bibinfo {author} {\bibfnamefont {M.~C.}\ \bibnamefont
  {Shapiro}}, \bibinfo {author} {\bibfnamefont {S.~C.}\ \bibnamefont {Riggs}},
  \bibinfo {author} {\bibfnamefont {M.~B.}\ \bibnamefont {Stone}}, \bibinfo
  {author} {\bibfnamefont {C.~R.}\ \bibnamefont {de~la Cruz}}, \bibinfo
  {author} {\bibfnamefont {S.}~\bibnamefont {Chi}}, \bibinfo {author}
  {\bibfnamefont {A.~A.}\ \bibnamefont {Podlesnyak}}, \ and\ \bibinfo {author}
  {\bibfnamefont {I.~R.}\ \bibnamefont {Fisher}},\ }\href {\doibase
  10.1103/PhysRevB.85.214434} {\bibfield  {journal} {\bibinfo  {journal} {Phys.
  Rev. B}\ }\textbf {\bibinfo {volume} {85}},\ \bibinfo {pages} {214434}
  (\bibinfo {year} {2012})}\BibitemShut {NoStop}%
\bibitem [{\citenamefont {Graf}\ \emph {et~al.}(2014)\citenamefont {Graf},
  \citenamefont {Disseler}, \citenamefont {Dhital}, \citenamefont {Hogan},
  \citenamefont {Bojko}, \citenamefont {Amato}, \citenamefont {Luetkens},
  \citenamefont {Baines}, \citenamefont {Margineda}, \citenamefont {Giblin},
  \citenamefont {Jura},\ and\ \citenamefont
  {Wilson}}]{intermediate-phase-Nd-Sm}%
  \BibitemOpen
  \bibfield  {author} {\bibinfo {author} {\bibfnamefont {M.~J.}\ \bibnamefont
  {Graf}}, \bibinfo {author} {\bibfnamefont {S.~M.}\ \bibnamefont {Disseler}},
  \bibinfo {author} {\bibfnamefont {C.}~\bibnamefont {Dhital}}, \bibinfo
  {author} {\bibfnamefont {T.}~\bibnamefont {Hogan}}, \bibinfo {author}
  {\bibfnamefont {M.}~\bibnamefont {Bojko}}, \bibinfo {author} {\bibfnamefont
  {A.}~\bibnamefont {Amato}}, \bibinfo {author} {\bibfnamefont
  {H.}~\bibnamefont {Luetkens}}, \bibinfo {author} {\bibfnamefont
  {C.}~\bibnamefont {Baines}}, \bibinfo {author} {\bibfnamefont
  {D.}~\bibnamefont {Margineda}}, \bibinfo {author} {\bibfnamefont {S.~R.}\
  \bibnamefont {Giblin}}, \bibinfo {author} {\bibfnamefont {M.}~\bibnamefont
  {Jura}}, \ and\ \bibinfo {author} {\bibfnamefont {S.~D.}\ \bibnamefont
  {Wilson}},\ }\href {http://stacks.iop.org/1742-6596/551/i=1/a=012020}
  {\bibfield  {journal} {\bibinfo  {journal} {Journal of Physics: Conference
  Series}\ }\textbf {\bibinfo {volume} {551}},\ \bibinfo {pages} {012020}
  (\bibinfo {year} {2014})}\BibitemShut {NoStop}%
\bibitem [{\citenamefont {Tomiyasu}\ \emph {et~al.}(2012)\citenamefont
  {Tomiyasu}, \citenamefont {Matsuhira}, \citenamefont {Iwasa}, \citenamefont
  {Watahiki}, \citenamefont {Takagi}, \citenamefont {Wakeshima}, \citenamefont
  {Hinatsu}, \citenamefont {Yokoyama}, \citenamefont {Ohoyama},\ and\
  \citenamefont {Yamada}}]{AIAO-Nd}%
  \BibitemOpen
  \bibfield  {author} {\bibinfo {author} {\bibfnamefont {K.}~\bibnamefont
  {Tomiyasu}}, \bibinfo {author} {\bibfnamefont {K.}~\bibnamefont {Matsuhira}},
  \bibinfo {author} {\bibfnamefont {K.}~\bibnamefont {Iwasa}}, \bibinfo
  {author} {\bibfnamefont {M.}~\bibnamefont {Watahiki}}, \bibinfo {author}
  {\bibfnamefont {S.}~\bibnamefont {Takagi}}, \bibinfo {author} {\bibfnamefont
  {M.}~\bibnamefont {Wakeshima}}, \bibinfo {author} {\bibfnamefont
  {Y.}~\bibnamefont {Hinatsu}}, \bibinfo {author} {\bibfnamefont
  {M.}~\bibnamefont {Yokoyama}}, \bibinfo {author} {\bibfnamefont
  {K.}~\bibnamefont {Ohoyama}}, \ and\ \bibinfo {author} {\bibfnamefont
  {K.}~\bibnamefont {Yamada}},\ }\href {\doibase 10.1143/JPSJ.81.034709}
  {\bibfield  {journal} {\bibinfo  {journal} {Journal of the Physical Society
  of Japan}\ }\textbf {\bibinfo {volume} {81}},\ \bibinfo {pages} {034709}
  (\bibinfo {year} {2012})}\BibitemShut {NoStop}%
\bibitem [{\citenamefont {Sagayama}\ \emph {et~al.}(2013)\citenamefont
  {Sagayama}, \citenamefont {Uematsu}, \citenamefont {Arima}, \citenamefont
  {Sugimoto}, \citenamefont {Ishikawa}, \citenamefont {O'Farrell},\ and\
  \citenamefont {Nakatsuji}}]{AIAO-Eu}%
  \BibitemOpen
  \bibfield  {author} {\bibinfo {author} {\bibfnamefont {H.}~\bibnamefont
  {Sagayama}}, \bibinfo {author} {\bibfnamefont {D.}~\bibnamefont {Uematsu}},
  \bibinfo {author} {\bibfnamefont {T.}~\bibnamefont {Arima}}, \bibinfo
  {author} {\bibfnamefont {K.}~\bibnamefont {Sugimoto}}, \bibinfo {author}
  {\bibfnamefont {J.~J.}\ \bibnamefont {Ishikawa}}, \bibinfo {author}
  {\bibfnamefont {E.}~\bibnamefont {O'Farrell}}, \ and\ \bibinfo {author}
  {\bibfnamefont {S.}~\bibnamefont {Nakatsuji}},\ }\href {\doibase
  10.1103/PhysRevB.87.100403} {\bibfield  {journal} {\bibinfo  {journal} {Phys.
  Rev. B}\ }\textbf {\bibinfo {volume} {87}},\ \bibinfo {pages} {100403}
  (\bibinfo {year} {2013})}\BibitemShut {NoStop}%
\bibitem [{\citenamefont {Disseler}(2014)}]{AIAO-Y-Eu-Nd}%
  \BibitemOpen
  \bibfield  {author} {\bibinfo {author} {\bibfnamefont {S.~M.}\ \bibnamefont
  {Disseler}},\ }\href {\doibase 10.1103/PhysRevB.89.140413} {\bibfield
  {journal} {\bibinfo  {journal} {Phys. Rev. B}\ }\textbf {\bibinfo {volume}
  {89}},\ \bibinfo {pages} {140413} (\bibinfo {year} {2014})}\BibitemShut
  {NoStop}%
\bibitem [{\citenamefont {Guo}\ \emph {et~al.}(2016)\citenamefont {Guo},
  \citenamefont {Ritter},\ and\ \citenamefont
  {Komarek}}]{Determination-AIAO-Nd2Ir2O7}%
  \BibitemOpen
  \bibfield  {author} {\bibinfo {author} {\bibfnamefont {H.}~\bibnamefont
  {Guo}}, \bibinfo {author} {\bibfnamefont {C.}~\bibnamefont {Ritter}}, \ and\
  \bibinfo {author} {\bibfnamefont {A.~C.}\ \bibnamefont {Komarek}},\ }\href
  {\doibase 10.1103/PhysRevB.94.161102} {\bibfield  {journal} {\bibinfo
  {journal} {Phys. Rev. B}\ }\textbf {\bibinfo {volume} {94}},\ \bibinfo
  {pages} {161102} (\bibinfo {year} {2016})}\BibitemShut {NoStop}%
\bibitem [{\citenamefont {Sushkov}\ \emph {et~al.}(2015)\citenamefont
  {Sushkov}, \citenamefont {Hofmann}, \citenamefont {Jenkins}, \citenamefont
  {Ishikawa}, \citenamefont {Nakatsuji}, \citenamefont {Das~Sarma},\ and\
  \citenamefont {Drew}}]{WSM-optical-Eu}%
  \BibitemOpen
  \bibfield  {author} {\bibinfo {author} {\bibfnamefont {A.~B.}\ \bibnamefont
  {Sushkov}}, \bibinfo {author} {\bibfnamefont {J.~B.}\ \bibnamefont
  {Hofmann}}, \bibinfo {author} {\bibfnamefont {G.~S.}\ \bibnamefont
  {Jenkins}}, \bibinfo {author} {\bibfnamefont {J.}~\bibnamefont {Ishikawa}},
  \bibinfo {author} {\bibfnamefont {S.}~\bibnamefont {Nakatsuji}}, \bibinfo
  {author} {\bibfnamefont {S.}~\bibnamefont {Das~Sarma}}, \ and\ \bibinfo
  {author} {\bibfnamefont {H.~D.}\ \bibnamefont {Drew}},\ }\href {\doibase
  10.1103/PhysRevB.92.241108} {\bibfield  {journal} {\bibinfo  {journal} {Phys.
  Rev. B}\ }\textbf {\bibinfo {volume} {92}},\ \bibinfo {pages} {241108}
  (\bibinfo {year} {2015})}\BibitemShut {NoStop}%
\bibitem [{\citenamefont {Nakayama}\ \emph {et~al.}(2016)\citenamefont
  {Nakayama}, \citenamefont {Kondo}, \citenamefont {Tian}, \citenamefont
  {Ishikawa}, \citenamefont {Halim}, \citenamefont {Bareille}, \citenamefont
  {Malaeb}, \citenamefont {Kuroda}, \citenamefont {Tomita}, \citenamefont
  {Ideta}, \citenamefont {Tanaka}, \citenamefont {Matsunami}, \citenamefont
  {Kimura}, \citenamefont {Inami}, \citenamefont {Ono}, \citenamefont
  {Kumigashira}, \citenamefont {Balents}, \citenamefont {Nakatsuji},\ and\
  \citenamefont {Shin}}]{WSM-ARPES-Nd}%
  \BibitemOpen
  \bibfield  {author} {\bibinfo {author} {\bibfnamefont {M.}~\bibnamefont
  {Nakayama}}, \bibinfo {author} {\bibfnamefont {T.}~\bibnamefont {Kondo}},
  \bibinfo {author} {\bibfnamefont {Z.}~\bibnamefont {Tian}}, \bibinfo {author}
  {\bibfnamefont {J.~J.}\ \bibnamefont {Ishikawa}}, \bibinfo {author}
  {\bibfnamefont {M.}~\bibnamefont {Halim}}, \bibinfo {author} {\bibfnamefont
  {C.}~\bibnamefont {Bareille}}, \bibinfo {author} {\bibfnamefont
  {W.}~\bibnamefont {Malaeb}}, \bibinfo {author} {\bibfnamefont
  {K.}~\bibnamefont {Kuroda}}, \bibinfo {author} {\bibfnamefont
  {T.}~\bibnamefont {Tomita}}, \bibinfo {author} {\bibfnamefont
  {S.}~\bibnamefont {Ideta}}, \bibinfo {author} {\bibfnamefont
  {K.}~\bibnamefont {Tanaka}}, \bibinfo {author} {\bibfnamefont
  {M.}~\bibnamefont {Matsunami}}, \bibinfo {author} {\bibfnamefont
  {S.}~\bibnamefont {Kimura}}, \bibinfo {author} {\bibfnamefont
  {N.}~\bibnamefont {Inami}}, \bibinfo {author} {\bibfnamefont
  {K.}~\bibnamefont {Ono}}, \bibinfo {author} {\bibfnamefont {H.}~\bibnamefont
  {Kumigashira}}, \bibinfo {author} {\bibfnamefont {L.}~\bibnamefont
  {Balents}}, \bibinfo {author} {\bibfnamefont {S.}~\bibnamefont {Nakatsuji}},
  \ and\ \bibinfo {author} {\bibfnamefont {S.}~\bibnamefont {Shin}},\
  }\href@noop {} {} (\bibinfo {year} {2016}),\ \Eprint
  {http://arxiv.org/abs/arXiv:1603.06095} {arXiv:1603.06095} \BibitemShut
  {NoStop}%
\bibitem [{\citenamefont {Shinaoka}\ \emph {et~al.}(2015)\citenamefont
  {Shinaoka}, \citenamefont {Hoshino}, \citenamefont {Troyer},\ and\
  \citenamefont {Werner}}]{lda-dmft-Y}%
  \BibitemOpen
  \bibfield  {author} {\bibinfo {author} {\bibfnamefont {H.}~\bibnamefont
  {Shinaoka}}, \bibinfo {author} {\bibfnamefont {S.}~\bibnamefont {Hoshino}},
  \bibinfo {author} {\bibfnamefont {M.}~\bibnamefont {Troyer}}, \ and\ \bibinfo
  {author} {\bibfnamefont {P.}~\bibnamefont {Werner}},\ }\href {\doibase
  10.1103/PhysRevLett.115.156401} {\bibfield  {journal} {\bibinfo  {journal}
  {Phys. Rev. Lett.}\ }\textbf {\bibinfo {volume} {115}},\ \bibinfo {pages}
  {156401} (\bibinfo {year} {2015})}\BibitemShut {NoStop}%
\bibitem [{\citenamefont {Zhang}\ \emph {et~al.}(2015)\citenamefont {Zhang},
  \citenamefont {Haule},\ and\ \citenamefont {Vanderbilt}}]{lda-dmft-R-dep}%
  \BibitemOpen
  \bibfield  {author} {\bibinfo {author} {\bibfnamefont {H.}~\bibnamefont
  {Zhang}}, \bibinfo {author} {\bibfnamefont {K.}~\bibnamefont {Haule}}, \ and\
  \bibinfo {author} {\bibfnamefont {D.}~\bibnamefont {Vanderbilt}},\
  }\href@noop {} {} (\bibinfo {year} {2015}),\ \Eprint
  {http://arxiv.org/abs/arXiv:1505.01203} {arXiv:1505.01203} \BibitemShut
  {NoStop}%
\bibitem [{\citenamefont {Kresse}\ and\ \citenamefont {Hafner}(1993)}]{vasp1}%
  \BibitemOpen
  \bibfield  {author} {\bibinfo {author} {\bibfnamefont {G.}~\bibnamefont
  {Kresse}}\ and\ \bibinfo {author} {\bibfnamefont {J.}~\bibnamefont
  {Hafner}},\ }\href@noop {} {\bibfield  {journal} {\bibinfo  {journal}
  {Physical Review B}\ }\textbf {\bibinfo {volume} {47}},\ \bibinfo {pages}
  {558} (\bibinfo {year} {1993})}\BibitemShut {NoStop}%
\bibitem [{\citenamefont {Kresse}\ and\ \citenamefont {Hafner}(1994)}]{vasp2}%
  \BibitemOpen
  \bibfield  {author} {\bibinfo {author} {\bibfnamefont {G.}~\bibnamefont
  {Kresse}}\ and\ \bibinfo {author} {\bibfnamefont {J.}~\bibnamefont
  {Hafner}},\ }\href@noop {} {\bibfield  {journal} {\bibinfo  {journal}
  {Physical Review B}\ }\textbf {\bibinfo {volume} {49}},\ \bibinfo {pages}
  {14251} (\bibinfo {year} {1994})}\BibitemShut {NoStop}%
\bibitem [{\citenamefont {Kresse}\ and\ \citenamefont
  {Furthm{\"u}ller}(1996{\natexlab{a}})}]{vasp3}%
  \BibitemOpen
  \bibfield  {author} {\bibinfo {author} {\bibfnamefont {G.}~\bibnamefont
  {Kresse}}\ and\ \bibinfo {author} {\bibfnamefont {J.}~\bibnamefont
  {Furthm{\"u}ller}},\ }\href@noop {} {\bibfield  {journal} {\bibinfo
  {journal} {Computational Materials Science}\ }\textbf {\bibinfo {volume}
  {6}},\ \bibinfo {pages} {15} (\bibinfo {year}
  {1996}{\natexlab{a}})}\BibitemShut {NoStop}%
\bibitem [{\citenamefont {Kresse}\ and\ \citenamefont
  {Furthm{\"u}ller}(1996{\natexlab{b}})}]{vasp4}%
  \BibitemOpen
  \bibfield  {author} {\bibinfo {author} {\bibfnamefont {G.}~\bibnamefont
  {Kresse}}\ and\ \bibinfo {author} {\bibfnamefont {J.}~\bibnamefont
  {Furthm{\"u}ller}},\ }\href@noop {} {\bibfield  {journal} {\bibinfo
  {journal} {Physical Review B}\ }\textbf {\bibinfo {volume} {54}},\ \bibinfo
  {pages} {11169} (\bibinfo {year} {1996}{\natexlab{b}})}\BibitemShut {NoStop}%
\bibitem [{\citenamefont {Kresse}\ and\ \citenamefont {Joubert}(1999)}]{vasp5}%
  \BibitemOpen
  \bibfield  {author} {\bibinfo {author} {\bibfnamefont {G.}~\bibnamefont
  {Kresse}}\ and\ \bibinfo {author} {\bibfnamefont {D.}~\bibnamefont
  {Joubert}},\ }\href@noop {} {\bibfield  {journal} {\bibinfo  {journal}
  {Physical Review B}\ }\textbf {\bibinfo {volume} {59}},\ \bibinfo {pages}
  {1758} (\bibinfo {year} {1999})}\BibitemShut {NoStop}%
\bibitem [{\citenamefont {Bl{\"o}chl}(1994)}]{PAW}%
  \BibitemOpen
  \bibfield  {author} {\bibinfo {author} {\bibfnamefont {P.~E.}\ \bibnamefont
  {Bl{\"o}chl}},\ }\href@noop {} {\bibfield  {journal} {\bibinfo  {journal}
  {Physical Review B}\ }\textbf {\bibinfo {volume} {50}},\ \bibinfo {pages}
  {17953} (\bibinfo {year} {1994})}\BibitemShut {NoStop}%
\bibitem [{\citenamefont {Perdew}\ \emph {et~al.}(1996)\citenamefont {Perdew},
  \citenamefont {Burke},\ and\ \citenamefont {Ernzerhof}}]{PBE}%
  \BibitemOpen
  \bibfield  {author} {\bibinfo {author} {\bibfnamefont {J.~P.}\ \bibnamefont
  {Perdew}}, \bibinfo {author} {\bibfnamefont {K.}~\bibnamefont {Burke}}, \
  and\ \bibinfo {author} {\bibfnamefont {M.}~\bibnamefont {Ernzerhof}},\
  }\href@noop {} {\bibfield  {journal} {\bibinfo  {journal} {Physical Review
  Letters}\ }\textbf {\bibinfo {volume} {77}},\ \bibinfo {pages} {3865}
  (\bibinfo {year} {1996})}\BibitemShut {NoStop}%
\bibitem [{\citenamefont {Marzari}\ and\ \citenamefont
  {Vanderbilt}(1997)}]{MLWF}%
  \BibitemOpen
  \bibfield  {author} {\bibinfo {author} {\bibfnamefont {N.}~\bibnamefont
  {Marzari}}\ and\ \bibinfo {author} {\bibfnamefont {D.}~\bibnamefont
  {Vanderbilt}},\ }\href@noop {} {\bibfield  {journal} {\bibinfo  {journal}
  {Physical Review B}\ }\textbf {\bibinfo {volume} {56}},\ \bibinfo {pages}
  {12847} (\bibinfo {year} {1997})}\BibitemShut {NoStop}%
\bibitem [{\citenamefont {Mostofi}\ \emph {et~al.}(2014)\citenamefont
  {Mostofi}, \citenamefont {Yates}, \citenamefont {Pizzi}, \citenamefont {Lee},
  \citenamefont {Souza}, \citenamefont {Vanderbilt},\ and\ \citenamefont
  {Marzari}}]{Wannier90}%
  \BibitemOpen
  \bibfield  {author} {\bibinfo {author} {\bibfnamefont {A.~A.}\ \bibnamefont
  {Mostofi}}, \bibinfo {author} {\bibfnamefont {J.~R.}\ \bibnamefont {Yates}},
  \bibinfo {author} {\bibfnamefont {G.}~\bibnamefont {Pizzi}}, \bibinfo
  {author} {\bibfnamefont {Y.-S.}\ \bibnamefont {Lee}}, \bibinfo {author}
  {\bibfnamefont {I.}~\bibnamefont {Souza}}, \bibinfo {author} {\bibfnamefont
  {D.}~\bibnamefont {Vanderbilt}}, \ and\ \bibinfo {author} {\bibfnamefont
  {N.}~\bibnamefont {Marzari}},\ }\href {\doibase
  http://dx.doi.org/10.1016/j.cpc.2014.05.003} {\bibfield  {journal} {\bibinfo
  {journal} {Computer Physics Communications}\ }\textbf {\bibinfo {volume}
  {185}},\ \bibinfo {pages} {2309 } (\bibinfo {year} {2014})}\BibitemShut
  {NoStop}%
\bibitem [{\citenamefont {Georges}\ \emph {et~al.}(1996)\citenamefont
  {Georges}, \citenamefont {Kotliar}, \citenamefont {Krauth},\ and\
  \citenamefont {Rozenberg}}]{dmft}%
  \BibitemOpen
  \bibfield  {author} {\bibinfo {author} {\bibfnamefont {A.}~\bibnamefont
  {Georges}}, \bibinfo {author} {\bibfnamefont {G.}~\bibnamefont {Kotliar}},
  \bibinfo {author} {\bibfnamefont {W.}~\bibnamefont {Krauth}}, \ and\ \bibinfo
  {author} {\bibfnamefont {M.~J.}\ \bibnamefont {Rozenberg}},\ }\href {\doibase
  10.1103/RevModPhys.68.13} {\bibfield  {journal} {\bibinfo  {journal} {Rev.
  Mod. Phys.}\ }\textbf {\bibinfo {volume} {68}},\ \bibinfo {pages} {13}
  (\bibinfo {year} {1996})}\BibitemShut {NoStop}%
\bibitem [{\citenamefont {Kotliar}\ \emph {et~al.}(2001)\citenamefont
  {Kotliar}, \citenamefont {Savrasov}, \citenamefont {P\'alsson},\ and\
  \citenamefont {Biroli}}]{CDMFT1}%
  \BibitemOpen
  \bibfield  {author} {\bibinfo {author} {\bibfnamefont {G.}~\bibnamefont
  {Kotliar}}, \bibinfo {author} {\bibfnamefont {S.~Y.}\ \bibnamefont
  {Savrasov}}, \bibinfo {author} {\bibfnamefont {G.}~\bibnamefont {P\'alsson}},
  \ and\ \bibinfo {author} {\bibfnamefont {G.}~\bibnamefont {Biroli}},\ }\href
  {\doibase 10.1103/PhysRevLett.87.186401} {\bibfield  {journal} {\bibinfo
  {journal} {Phys. Rev. Lett.}\ }\textbf {\bibinfo {volume} {87}},\ \bibinfo
  {pages} {186401} (\bibinfo {year} {2001})}\BibitemShut {NoStop}%
\bibitem [{\citenamefont {Bolech}\ \emph {et~al.}(2003)\citenamefont {Bolech},
  \citenamefont {Kancharla},\ and\ \citenamefont {Kotliar}}]{CDMFT2}%
  \BibitemOpen
  \bibfield  {author} {\bibinfo {author} {\bibfnamefont {C.~J.}\ \bibnamefont
  {Bolech}}, \bibinfo {author} {\bibfnamefont {S.~S.}\ \bibnamefont
  {Kancharla}}, \ and\ \bibinfo {author} {\bibfnamefont {G.}~\bibnamefont
  {Kotliar}},\ }\href {\doibase 10.1103/PhysRevB.67.075110} {\bibfield
  {journal} {\bibinfo  {journal} {Phys. Rev. B}\ }\textbf {\bibinfo {volume}
  {67}},\ \bibinfo {pages} {075110} (\bibinfo {year} {2003})}\BibitemShut
  {NoStop}%
\bibitem [{\citenamefont {Maier}\ \emph {et~al.}(2005)\citenamefont {Maier},
  \citenamefont {Jarrell}, \citenamefont {Pruschke},\ and\ \citenamefont
  {Hettler}}]{cluster}%
  \BibitemOpen
  \bibfield  {author} {\bibinfo {author} {\bibfnamefont {T.}~\bibnamefont
  {Maier}}, \bibinfo {author} {\bibfnamefont {M.}~\bibnamefont {Jarrell}},
  \bibinfo {author} {\bibfnamefont {T.}~\bibnamefont {Pruschke}}, \ and\
  \bibinfo {author} {\bibfnamefont {M.~H.}\ \bibnamefont {Hettler}},\ }\href
  {\doibase 10.1103/RevModPhys.77.1027} {\bibfield  {journal} {\bibinfo
  {journal} {Rev. Mod. Phys.}\ }\textbf {\bibinfo {volume} {77}},\ \bibinfo
  {pages} {1027} (\bibinfo {year} {2005})}\BibitemShut {NoStop}%
\bibitem [{\citenamefont {Koch}\ \emph {et~al.}(2008)\citenamefont {Koch},
  \citenamefont {Sangiovanni},\ and\ \citenamefont
  {Gunnarsson}}]{group-theory}%
  \BibitemOpen
  \bibfield  {author} {\bibinfo {author} {\bibfnamefont {E.}~\bibnamefont
  {Koch}}, \bibinfo {author} {\bibfnamefont {G.}~\bibnamefont {Sangiovanni}}, \
  and\ \bibinfo {author} {\bibfnamefont {O.}~\bibnamefont {Gunnarsson}},\
  }\href {\doibase 10.1103/PhysRevB.78.115102} {\bibfield  {journal} {\bibinfo
  {journal} {Phys. Rev. B}\ }\textbf {\bibinfo {volume} {78}},\ \bibinfo
  {pages} {115102} (\bibinfo {year} {2008})}\BibitemShut {NoStop}%
\bibitem [{\citenamefont {Caffarel}\ and\ \citenamefont {Krauth}(1994)}]{ED}%
  \BibitemOpen
  \bibfield  {author} {\bibinfo {author} {\bibfnamefont {M.}~\bibnamefont
  {Caffarel}}\ and\ \bibinfo {author} {\bibfnamefont {W.}~\bibnamefont
  {Krauth}},\ }\href {\doibase 10.1103/PhysRevLett.72.1545} {\bibfield
  {journal} {\bibinfo  {journal} {Phys. Rev. Lett.}\ }\textbf {\bibinfo
  {volume} {72}},\ \bibinfo {pages} {1545} (\bibinfo {year}
  {1994})}\BibitemShut {NoStop}%
\bibitem [{\citenamefont {Go}\ and\ \citenamefont {Jeon}(2009)}]{cdmft-model}%
  \BibitemOpen
  \bibfield  {author} {\bibinfo {author} {\bibfnamefont {A.}~\bibnamefont
  {Go}}\ and\ \bibinfo {author} {\bibfnamefont {G.~S.}\ \bibnamefont {Jeon}},\
  }\href {http://stacks.iop.org/0953-8984/21/i=48/a=485602} {\bibfield
  {journal} {\bibinfo  {journal} {Journal of Physics: Condensed Matter}\
  }\textbf {\bibinfo {volume} {21}},\ \bibinfo {pages} {485602} (\bibinfo
  {year} {2009})}\BibitemShut {NoStop}%
\bibitem [{\citenamefont {Witczak-Krempa}\ \emph {et~al.}(2014)\citenamefont
  {Witczak-Krempa}, \citenamefont {Chen}, \citenamefont {Kim},\ and\
  \citenamefont {Balents}}]{review}%
  \BibitemOpen
  \bibfield  {author} {\bibinfo {author} {\bibfnamefont {W.}~\bibnamefont
  {Witczak-Krempa}}, \bibinfo {author} {\bibfnamefont {G.}~\bibnamefont
  {Chen}}, \bibinfo {author} {\bibfnamefont {Y.~B.}\ \bibnamefont {Kim}}, \
  and\ \bibinfo {author} {\bibfnamefont {L.}~\bibnamefont {Balents}},\ }\href
  {\doibase 10.1146/annurev-conmatphys-020911-125138} {\bibfield  {journal}
  {\bibinfo  {journal} {Annual Review of Condensed Matter Physics}\ }\textbf
  {\bibinfo {volume} {5}},\ \bibinfo {pages} {57} (\bibinfo {year}
  {2014})}\BibitemShut {NoStop}%
\bibitem [{\citenamefont {Liu}\ \emph {et~al.}(2014)\citenamefont {Liu},
  \citenamefont {Tong}, \citenamefont {Ling}, \citenamefont {Zhang},
  \citenamefont {Zhang}, \citenamefont {Zhang}, \citenamefont {Pi},
  \citenamefont {Zhang},\ and\ \citenamefont {Zhang}}]{transport-Y}%
  \BibitemOpen
  \bibfield  {author} {\bibinfo {author} {\bibfnamefont {H.}~\bibnamefont
  {Liu}}, \bibinfo {author} {\bibfnamefont {W.}~\bibnamefont {Tong}}, \bibinfo
  {author} {\bibfnamefont {L.}~\bibnamefont {Ling}}, \bibinfo {author}
  {\bibfnamefont {S.}~\bibnamefont {Zhang}}, \bibinfo {author} {\bibfnamefont
  {R.}~\bibnamefont {Zhang}}, \bibinfo {author} {\bibfnamefont
  {L.}~\bibnamefont {Zhang}}, \bibinfo {author} {\bibfnamefont
  {L.}~\bibnamefont {Pi}}, \bibinfo {author} {\bibfnamefont {C.}~\bibnamefont
  {Zhang}}, \ and\ \bibinfo {author} {\bibfnamefont {Y.}~\bibnamefont
  {Zhang}},\ }\href {\doibase http://dx.doi.org/10.1016/j.ssc.2013.11.004}
  {\bibfield  {journal} {\bibinfo  {journal} {Solid State Communications}\
  }\textbf {\bibinfo {volume} {179}},\ \bibinfo {pages} {1 } (\bibinfo {year}
  {2014})}\BibitemShut {NoStop}%
\bibitem [{\citenamefont {Takatsu}\ \emph {et~al.}(2014)\citenamefont
  {Takatsu}, \citenamefont {Watanabe}, \citenamefont {Goto},\ and\
  \citenamefont {Kadowaki}}]{transport-Eu}%
  \BibitemOpen
  \bibfield  {author} {\bibinfo {author} {\bibfnamefont {H.}~\bibnamefont
  {Takatsu}}, \bibinfo {author} {\bibfnamefont {K.}~\bibnamefont {Watanabe}},
  \bibinfo {author} {\bibfnamefont {K.}~\bibnamefont {Goto}}, \ and\ \bibinfo
  {author} {\bibfnamefont {H.}~\bibnamefont {Kadowaki}},\ }\href {\doibase
  10.1103/PhysRevB.90.235110} {\bibfield  {journal} {\bibinfo  {journal} {Phys.
  Rev. B}\ }\textbf {\bibinfo {volume} {90}},\ \bibinfo {pages} {235110}
  (\bibinfo {year} {2014})}\BibitemShut {NoStop}%
\bibitem [{\citenamefont {Park}\ \emph {et~al.}(2008)\citenamefont {Park},
  \citenamefont {Haule},\ and\ \citenamefont {Kotliar}}]{Park08}%
  \BibitemOpen
  \bibfield  {author} {\bibinfo {author} {\bibfnamefont {H.}~\bibnamefont
  {Park}}, \bibinfo {author} {\bibfnamefont {K.}~\bibnamefont {Haule}}, \ and\
  \bibinfo {author} {\bibfnamefont {G.}~\bibnamefont {Kotliar}},\ }\href@noop
  {} {\bibfield  {journal} {\bibinfo  {journal} {Phys. Rev. Lett.}\ }\textbf
  {\bibinfo {volume} {101}},\ \bibinfo {pages} {186403} (\bibinfo {year}
  {2008})}\BibitemShut {NoStop}%
\bibitem [{\citenamefont {Gull}\ \emph {et~al.}(2008)\citenamefont {Gull},
  \citenamefont {Werner}, \citenamefont {Wang}, \citenamefont {Troyer},\ and\
  \citenamefont {Millis}}]{Gull08}%
  \BibitemOpen
  \bibfield  {author} {\bibinfo {author} {\bibfnamefont {E.}~\bibnamefont
  {Gull}}, \bibinfo {author} {\bibfnamefont {P.}~\bibnamefont {Werner}},
  \bibinfo {author} {\bibfnamefont {X.}~\bibnamefont {Wang}}, \bibinfo {author}
  {\bibfnamefont {M.}~\bibnamefont {Troyer}}, \ and\ \bibinfo {author}
  {\bibfnamefont {A.~J.}\ \bibnamefont {Millis}},\ }\href {\doibase
  10.1209/0295-5075/84/37009} {\bibfield  {journal} {\bibinfo  {journal}
  {Europhys. Lett.}\ }\textbf {\bibinfo {volume} {84}},\ \bibinfo {pages}
  {37009 (6pp)} (\bibinfo {year} {2008})}\BibitemShut {NoStop}%
\bibitem [{\citenamefont {Koster}(1963)}]{book}%
  \BibitemOpen
  \bibfield  {author} {\bibinfo {author} {\bibfnamefont {G.~F.}\ \bibnamefont
  {Koster}},\ }\href@noop {} {\emph {\bibinfo {title} {Properties of the
  thirty-two point groups}}}\ (\bibinfo  {publisher} {Cambridge, Mass., M.I.T.
  Press},\ \bibinfo {year} {1963})\BibitemShut {NoStop}%
\end{thebibliography}%

\end{document}